\documentclass[a4paper,fleqn,usenatbib]{mnras}

\usepackage[T1]{fontenc}
\usepackage{ae,aecompl}

\usepackage{graphicx}	
\usepackage{amsmath}	
\usepackage{amssymb}	
\usepackage{comment}
\usepackage{color}
\usepackage{enumitem}

\title[]{From hydrodynamics to \textit{N}-body simulations of star clusters: mergers and rotation}

\author[A. Ballone et al.]{Alessandro Ballone$^{1,2,3}$, Stefano Torniamenti$^{1,2,3}$,   Michela Mapelli$^{1,2,3}$,
\newauthor Ugo N. Di Carlo$^{4,1,2}$, Mario Spera$^{1,2,5,6}$, Sara Rastello$^{1,2}$, Nicola Gaspari$^{1,7}$,
\newauthor Giuliano Iorio$^{1,2}$
\\
$^{1}$Physics and Astronomy Department Galileo Galilei, University of Padova, Vicolo dell'Osservatorio 3, I--35122, Padova, Italy\\
$^{2}$INFN - Padova, Via Marzolo 8, I--35131 Padova, Italy\\
$^{3}$INAF - Osservatorio Astronomico di Padova, Vicolo dell'Osservatorio 5, I-35122 Padova, Italy\\
$^{4}$Dipartimento di Scienza e Alta Tecnologia, University of Insubria, Via Valleggio 11, I-22100 Como, Italy\\
$^{5}$Center for Interdisciplinary Exploration and Research in Astrophysics (CIERA), Evanston, IL 60208, USA\\
$^{6}$Department of Physics \& Astronomy, Northwestern University, Evanston, IL 60208, USA\\
$^{7}$Department of Astrophysics/IMAPP, Radboud University, P O Box 9010, NL-6500 GL Nijmegen, The Netherlands\\}

\date{Accepted XXX. Received YYY; in original form ZZZ}

\pubyear{2018}

\begin{document}
\label{firstpage}
\pagerange{\pageref{firstpage}--\pageref{lastpage}}
\maketitle

\begin{abstract}

We present a new method to obtain more realistic initial conditions for \textit{N}-body simulations of young star clusters. 
We start from the outputs of hydrodynamical simulations of molecular cloud collapse, in which star formation is modelled with sink particles. In our approach, we instantaneously remove gas from these hydrodynamical simulation outputs to mock the end of the gas-embedded phase, induced by stellar feedback. We then enforce a realistic initial mass function by splitting or joining the sink particles based on their mass and position.
Such initial conditions contain more consistent information on the spatial distribution and the kinematical and dynamical states of young star clusters, which are fundamental to properly study these systems. For example, by applying our method to a set of previously run hydrodynamical simulations, we found that the early evolution of young star clusters is affected by gas removal and by the early dry merging of sub-structures. This early evolution can either quickly erase the rotation acquired by our (sub-)clusters in their embedded phase or ``fuel'' it by feeding of angular momentum by sub-structure mergers, before two-body relaxation acts on longer timescales. 
\end{abstract}




\section{Introduction}\label{intro}

More and more observations and theoretical studies suggest that the properties of star clusters can be strongly affected by their formation process, through the collapse of turbulent clouds of molecular gas. 

For example, young massive clusters and open clusters show strong indications of sub-structures and fractality \citep[e.g.,][]{Cartwright04, Sanchez09, Parker12, Kuhn19, Cantat-Gaudin19}, traces of ongoing dispersal \citep[e.g.,][]{Kuhn19,Cantat-Gaudin19}, possibly led by sudden gas expulsion \citep[e.g.,][]{Tutukov78,Lada84,Geyer01, Baumgardt07, Pang20}, signatures of rotation \citep[e.g.,][]{Henault-Brunet12} and complex kinematics of sub-clumps \citep[e.g.,][]{Kuhn19,Cantat-Gaudin19}. 

Globular clusters are much older systems, where we would expect secular evolution to have washed out all the imprints of their formation process. Nonetheless, we still have observational and theoretical indications of the impact of their birth on some of their properties. For example, signatures of rotation are also found for these much older systems \citep[e.g.,][]{vanLeeuwen00,Pancino07,Bianchini13, Fabricius14, Kamann18, Bianchini18}. Moreover, though the stars in these systems seem to be roughly coheval, we observe multiple populations with slightly different chemical properties, kinematics and locations \citep[see the recent review by][and references therein]{Gratton19}. The formation of multiple populations is still an open question, but suggests that the formation of globular clusters might have occurred from a non-monolithic and perfectly simultaneous collapse of their parent molecular cloud \citep{Perets14,Mastrobuono-Battisti16,Gavagnin16,Bekki16,Mastrobuono-Battisti19}.

Much theoretical work has been done by means of hydrodynamical simulations, focusing on the process of star formation, often involving complex physics, such as accreting sink particles, stellar feedback, magnetic fields, chemistry \citep[e.g.,][]{Padoan11, Krumholz12, Dale15, Geen16, Seifried17, Nakauchi18, Zamora-Aviles19, Lee19}.
On the one hand, these studies require high resolution, needed to properly resolve the hydrodynamics of  gas collapse and to be able to sample the low mass part of the initial stellar mass function. Furthermore, the need to include complex physics (e.g., gas cooling, radiation treatment, evolution of chemical networks, etc.) often leads to a tremendous slow-down of the computation. So, these simulations often focus on the collapse of molecular clouds with mass $\lesssim 10^3 M_{\odot}$, relatively small compared to observed giant molecular clouds (which have masses $> 10^4 M_{\odot}$), and they are evolved for relatively short times, of the order of few free-fall times of the cloud.
On the other hand, hydrodynamical simulations of star cluster formation usually do not include direct \textit{N}-body algorithms \citep{Aarseth10}. As a result, the dynamics of the newly formed stars is in most cases integrated following a ``collisionless'' approach (i.e., without directly calculating the inter-particle forces, but adopting, for example, tree algorithms),
while young star clusters are collisional systems, i.e. their two-body relaxation timescale is much shorter than their lifetime \citep{Spitzer87, Portegies-Zwart10}.

Direct \textit{N}-body simulations are usually adopted to integrate the collisional dynamics of gas-free star clusters. In most cases, idealized initial conditions (such as Plummer or King models) are used. In some recent studies, fractal initial conditions are adopted to mimic the initial clumpiness of star clusters \citep[e.g.,][]{Goodwin04, Schmeja06,Allison10,Kuepper11,Parker14,DiCarlo19,Daffern-Powell20}.

A new approach to bridge hydrodynamics of gas and collisional dynamics of stars is needed: hydrodynamical simulations can potentially be used to provide more realistic initial conditions for direct \textit{N}-body simulations. This is possible, since the gas in which the newly formed star cluster is embedded is expected to be almost instantaneously expelled by feedback (radiation, winds and, most of all, supernova explosions) from the young most massive stars \citep[e.g.,][]{Tutukov78,Geyer01,Baumgardt07,Farias15}. From that moment on, we expect the evolution of the newly born stellar system to be mainly driven by gravitational dynamics. Very few attempts focused on this task. 

\citet{Hurley08} were the first who re-simulated the output of a simulation run with smoothed particle hydrodynamics (SPH) with a direct \textit{N}-body integrator, to study the evolution of stellar clusters in the galactic potential. 

\citet{Moeckel10} used the sink particles produced in a high-resolution SPH simulation by \citet{Bate09} as initial conditions for a simulation with the direct \textit{N}-body code {\sc NBODY6} \citep{Aarseth03}, focusing particularly on the evolution of binary and multiple stellar systems. The sink particles, in this case, had an almost realistic mass function down to very low stellar masses, comparable to a \citet{Kroupa01} or \citet{Chabrier03} initial mass function (IMF). With the same technique, \citet{Moeckel12} studied the dispersal of young stellar systems, after sudden gas removal, by starting from sink particles formed in the smoothed-particle-hydrodynamics (SPH) simulations of a larger ($10^4 M_{\odot}$) molecular cloud by \citet{Bonnell08}. In this case, however, the sink particles produced in their hydrodynamical simulation tended to have a shallower high-mass slope of their mass function, compared to observed ones \citep{Maschberger10}. Similarly, \citet{Parker13} re-simulated the dynamical evolution of the sinks formed in the highest resolution simulations of molecular clouds (with masses $1-3 \times 10^4 M_{\odot}$) of \citet{Dale12} and \citet{Dale13}, to test the impact of photoionization feedback on the initial conditions and early evolution of star clusters. The \textit{N}-body runs were performed with the {\sc starlab} package, coupling the 4th-order Hermite integrator {\sc kira} to the {\sc SeBa} code, to follow the single/binary stellar evolution \citep[e.g.,][]{Portegies-Zwart96, Portegies-Zwart01}.

A slightly different approach was followed by \citet{Fujii15a} \citep[see also][]{Fujii15b,Fujii16,Fujii19} to study the mass function and kinematics of sub-clumps in young stellar clusters produced by giant molecular clouds. 
In their approach, the hydrodynamical simulations were run without sink particles up to roughly the free-fall time of the clouds and then the densest gas particles were converted into stars, by assuming a certain star formation efficiency. In this conversion, each ``star-forming'' gas particle was substituted by a new star particle which directly inherited the position and velocity of its ``parent'' gas particle, while its mass was selected randomly from a \citet{Salpeter55} mass function between 0.3 and 100 M$_{\odot}$. This method conserved the mass globally (though not locally), since the mass resolution of the hydrodynamical simulation was 1 M$_{\odot}$, equal to the average mass of the latter mass function.

Recently, \citet{Wall19} and \citet{Cournoyer-Cloutier20} simulated the formation and dynamical evolution of a star cluster through the {\sc amuse} framework \citep{Pelupessy13}. For these simulations, the authors accurately and simultaneously computed hydrodynamics, with the Eulerian adaptive-mesh refinement (AMR) code {\sc FLASH} \citep{Fryxell00}, and stellar dynamics, with the direct \textit{N}-body code {\sc ph4} \citep{McMillan12}.\footnote{In these simulations, the authors also had a quite realistic treatment of gas thermodynamics, the further coupling with a stellar evolution code {\sc SeBa} \citep{Portegies-Zwart96}, the inclusion of stellar feedback on the gas and a method to specifically resolve binary/multiple evolution.} In this work, a realistic stellar mass function was obtained, by forcing the sink particles forming on the fly to follow a \citet{Kroupa01} mass function between 0.08 and 150 M$_{\odot}$. The one presented in \cite{Wall19} and \citet{Cournoyer-Cloutier20} is perhaps the most self-consistent attempt to combine hydrodynamics and collisional dynamics, but is certainly very computationally expensive; hence, it does not easily allow to explore the relevant parameter space for star cluster formation.

In this work, we build on these previous attempts. In section \ref{methods}, we  present our new method for getting more realistic initial conditions for \textit{N}-body simulations. The major advance of this new approach consists in starting from the properties of sink particles in hydrodynamical simulations of star formation, but using them to construct star clusters with  realistic mass functions. Having realistic mass functions is crucial to study many different topics. In Section \ref{tests}, we present an analysis of the ``performance'' of the method. In Section \ref{results}, we show the results of the re-simulation of few stellar clusters with direct \textit{N}-body, focusing on the early evolution of our sub-clusters and on the merging and the rotation of sub-structures.

\begin{figure}
\begin{center}
\includegraphics[scale=0.7,trim={0 0.4cm 0 0.4cm},clip]{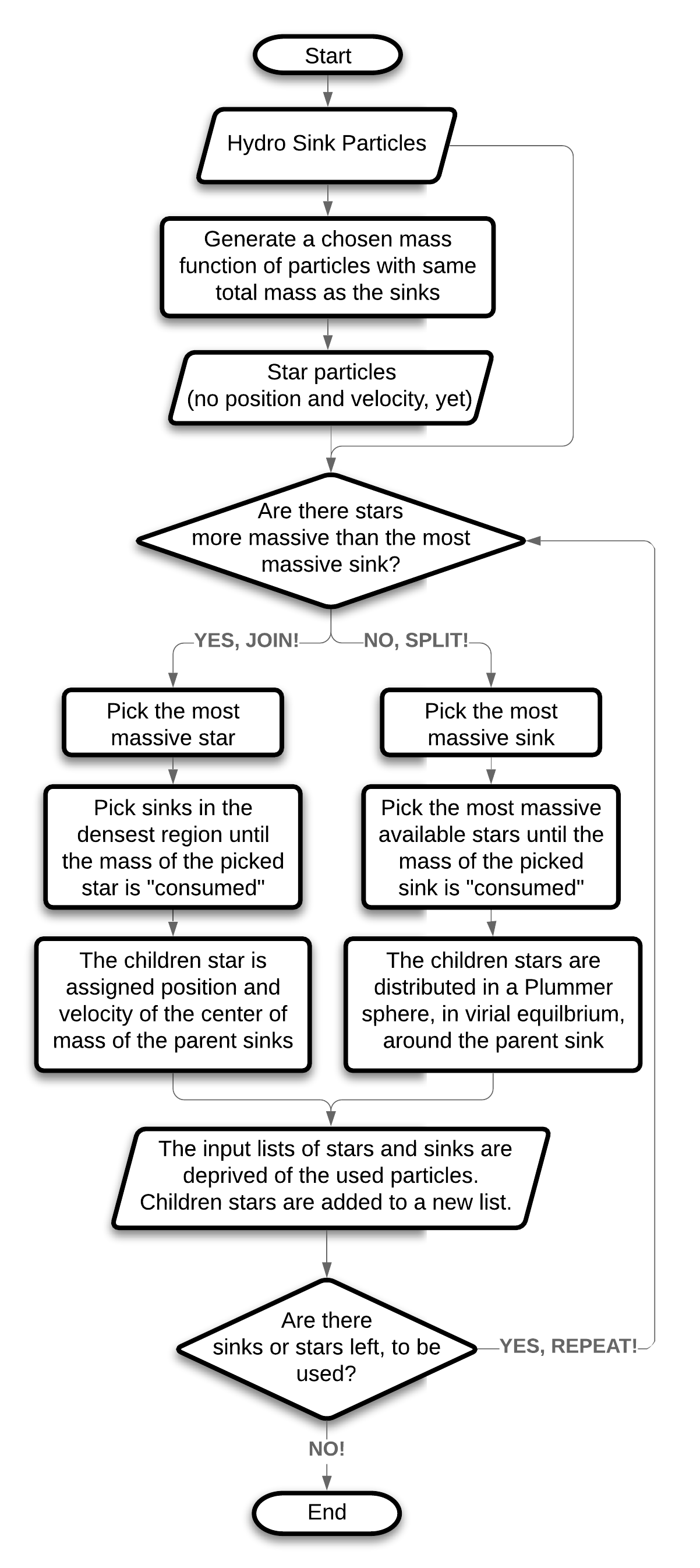}
\caption{Flow-chart for our joining/splitting algorithm to obtain a realistic distribution of stars from sink particles of hydrodynamical simulations.
}\label{flow_chart}
\end{center}
\end{figure}

\begin{figure}
\begin{center}
\includegraphics[scale=0.56]{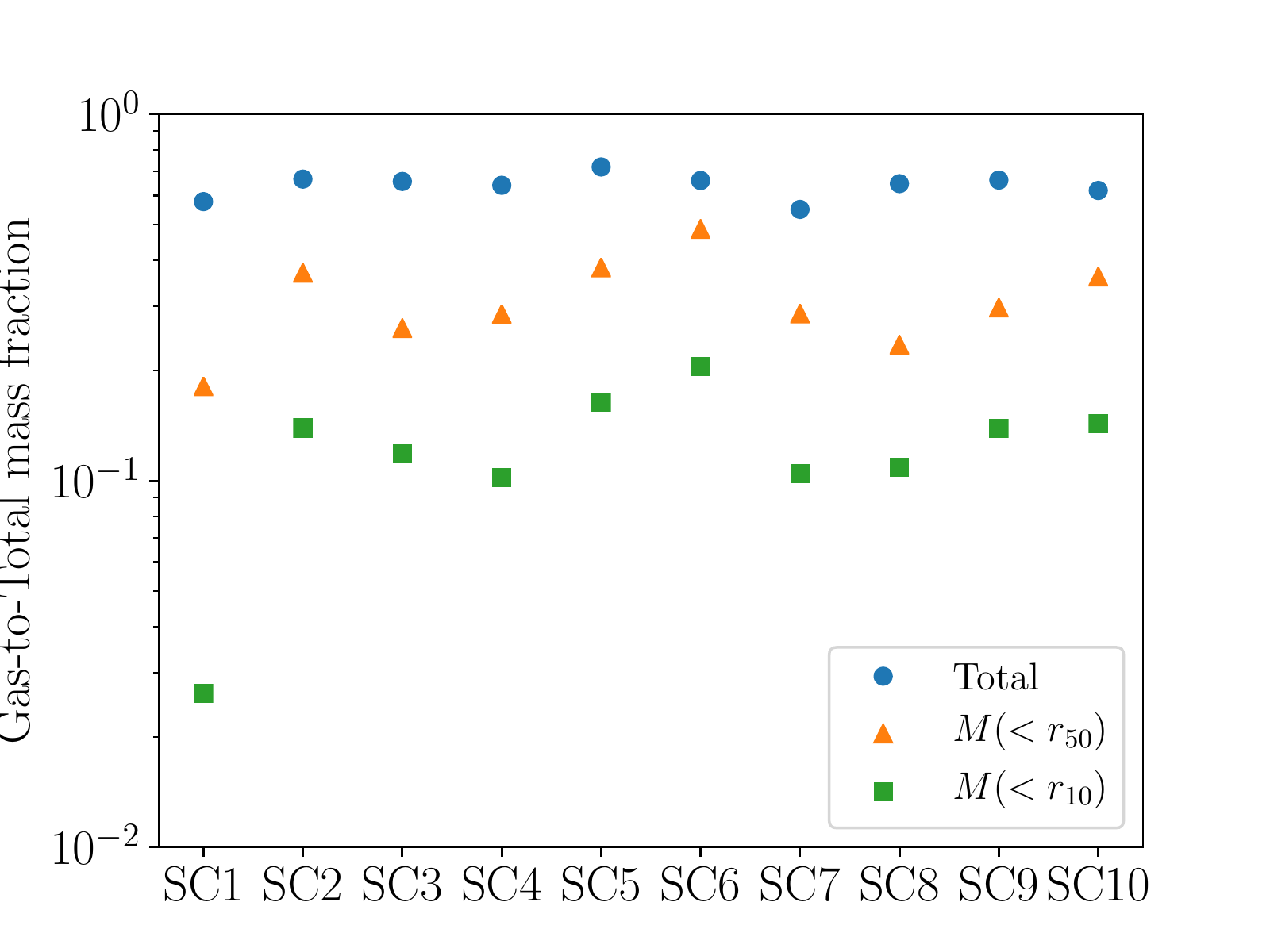}
\caption{Gas-to-total mass fractions for our set, calculated for the total cluster (blue circles), within $r_{50}$ (orange triangles) and within $r_{10}$ (green squares).
}\label{mass}
\end{center}
\end{figure}

\begin{figure*}
\begin{center}
\includegraphics[scale=0.75,trim={0 2cm 0 2cm},clip]{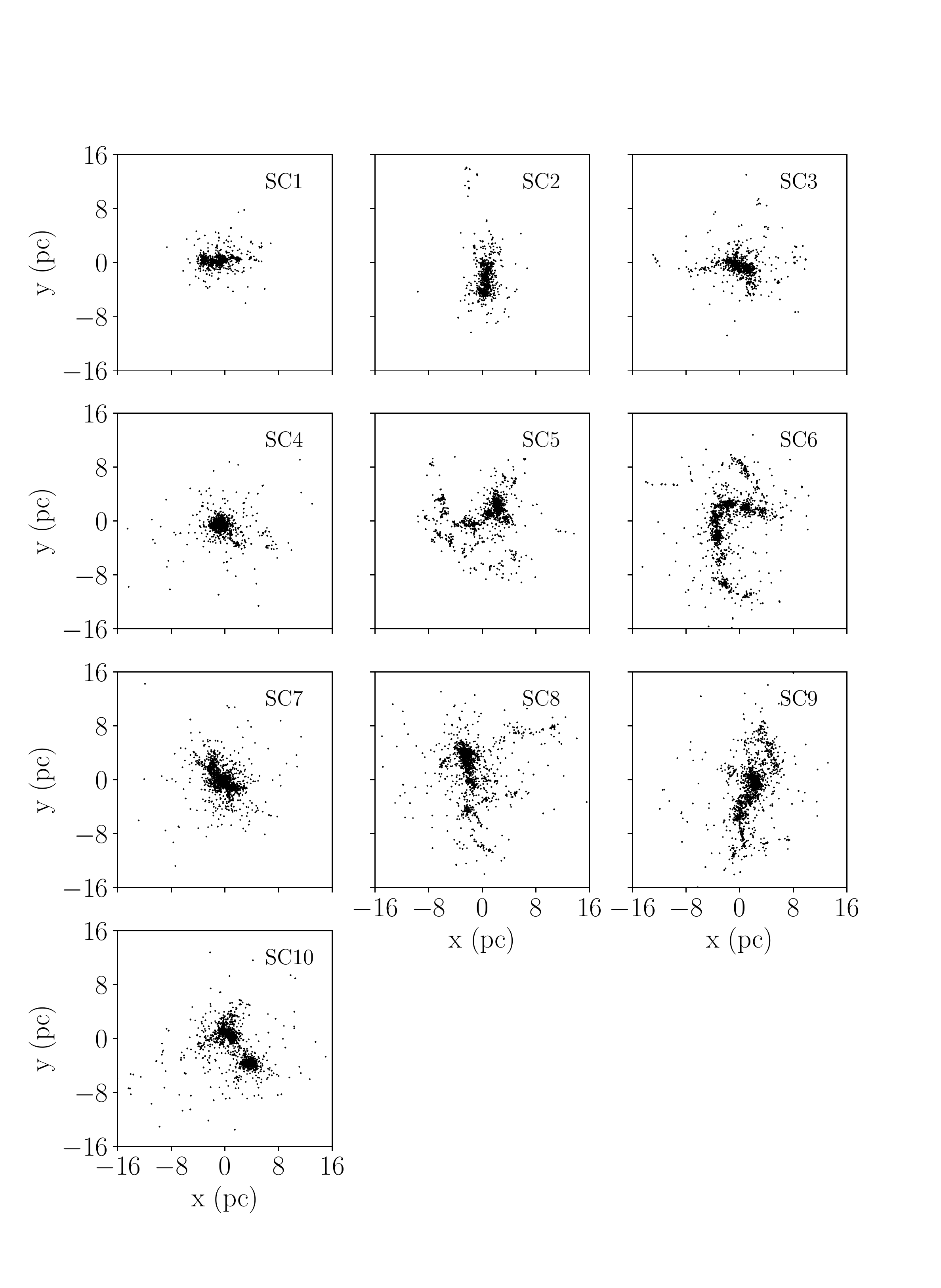}
\caption{Spatial distribution of our star clusters, after applying our joining/splitting algorithm. These distributions correspond to the initial conditions of our \textit{N}-body simulations.
}\label{map_stars}
\end{center}
\end{figure*}

\section{Methods}\label{methods}

\subsection{The joining/splitting algorithm} 

Our method to generate more realistic initial conditions for \textit{N}-body simulations can be applied to any hydrodynamical simulations run with sink particles. The main idea behind it is based on joining or splitting the sink particles in mass, in order to recover a new chosen, more realistic mass function for the ``children'' stars. In fact, obtaining a realistic IMF for sink particles in hydrodynamical simulations is not an easy task, since the latter might strongly be affected by resolution and by the adopted sink particle algorithm.

In particular, our sink joining/splitting algorithm consists of the following two major steps.

\begin{enumerate}[label=(\alph*),align=left]

\item A list of ``children'' star particles is generated, with mass distribution following a chosen mass function and total mass equal to the total mass of the sink particles.\\

\item Sink particles are joined (b1) or split (b2) to produce  children stars; i.e., the mass of few sink particles is summed up to reach the mass of a children star or the mass of a sink particle is split into few children stars. The children stars then inherit the position $\mathbf{x}$ and velocity $\mathbf{v}$ of their parent sinks.

\end{enumerate}

In the following, we detail these steps (see also the flow chart in Fig. \ref{flow_chart}).

\begin{enumerate}[label=(\alph*),align=left]
\item In the first step, after choosing a certain mass function \citep[e.g., in the following we adopt a Kroupa mass function:][]{Kroupa01}, we randomly generate masses for the children stars so that they follow the chosen probability distribution function in mass. We repeat the generation of stars until the total mass of the generated sample of particles is equal to the total mass of the hydrodynamical sinks, within 0.2 M$_{\odot}$.

\item In the second step, we cycle over the hydro sinks and the children stars, until one of the two lists of particles is emptied. In each step of the cycle we decide whether to apply a \textbf{\textit{joining}} of sinks to recover a star or a \textbf{\textit{splitting}} of sinks into several stars.

(b1) If there is at least one star that is more massive than the most massive sink, we apply the \textbf{\textit{joining branch}} of our algorithm. In this case, we check the sink particles in their position space and calculate the densest region of sink particles, by calculating the total mass of particles contained in a sphere with radius 0.1 pc, centered at the position of each sink particle. After finding the densest sphere, we sum up the mass of $n$ random sinks contained therein, until the mass of the most massive star is reached. We then assign this (new, but slightly different)\footnote{We prefer to update the mass of the star to this new value, in order to better enforce mass conservation. Nonetheless, we find that the new mass is always different from less than few percent of the mass of the star generated in step (a).} ``joining'' mass to the children star, along with a position and velocity in phase space, obtained by conserving the center of mass of the employed sinks. However, we re-scale the velocity modulus of the new star, so to conserve the total kinetic energy of the parent sinks. This is a somewhat arbitrary choice. However, by testing both pure momentum and kinetic energy conservation, we can affirm that there is very little difference in terms of the total kinetic energy and virial ratios of the resulting star clusters. The used sinks and the children star are then removed from the lists and the latter is saved as a newborn star. 

(b2) If the most massive sink is more massive than any star, we apply the \textbf{\textit{splitting branch}} of our algorithm. In this case, we take the most massive sink and start subtracting to it the mass of the most massive star available. We repeat this operation until the remaining mass is smaller than 0.1 M$_{\odot}$, storing the employed stars in a ``mini cluster'' sublist. At this point the leftover mass is reassigned to the sink closest in space to the ``consumed'' one, so to enforce local and total mass conservation. The ``mini cluster'' stars are then distributed around the position and velocity of their parent sink, following a Plummer distribution in virial equilibrium, with radius\footnote{This step is performed using the {\sc new\_plummer\_model} module in {\sc amuse} \citep{Pelupessy13}}  $r_{plum}$ (see Sec. \ref{rplum_sec} for further details). At the end of this further operation, the used sink and the children stars are removed from their original lists and the latter are added to the final list of newborn stars.
\end{enumerate}

\begin{table}
\caption{Properties of the star clusters, obtained through our joining/splitting algorithm, used as initial conditions of the \textit{N}-body simulations performed for this study.}
\label{tab1}
\centering  
\begin{tabular}{l l l l l}
\hline
Name & Total mass & Number of stars & r$_{50}$ & r$_{10}$ \\
 & (M$_{\odot}$) & & (pc) & (pc) \\
\hline
SC1 & 4218.9 & 6254 & 1.21 & 0.06 \\
SC2 & 6687.0 & 9747 & 2.11 & 0.17 \\
SC3 & 10323.0 & 15152 & 1.49 & 0.16 \\
SC4 & 14391.3 & 21060 & 1.93 & 0.46 \\
SC5 & 14086.5 & 20203 & 3.49 & 0.34 \\
SC6 & 20412.2 & 29563 & 6.77 & 0.71 \\
SC7 & 31465.8 & 45462 & 3.36 & 0.19 \\
SC8 & 28250.0 & 40849 & 3.16 & 0.21 \\
SC9 & 30460.9 & 44229 & 4.60 & 0.41 \\
SC10 & 38018.5 & 55295 & 4.63 & 0.33 \\
\hline
\end{tabular}
\end{table}

\subsection{Hydrodynamical simulations}\label{hydro}

We applied our method to the 10 hydrodynamical simulations of turbulent molecular clouds presented in \citet{Ballone20}, performed with the SPH code {\sc GASOLINE2} \citep{Wadsley04, Wadsley17}.
The clouds have an initial uniform density and temperature of 250 $\mathrm{cm^{-3}}$ and 10 K, respectively. The gas particles are distributed in a sphere with total mass in the range $[10^4-10^5] \; M_{\odot}$. All the clouds are initially turbulent, so to be in a marginally bound state (i.e., their virial ratio $\alpha_{vir}=T/|V|=1$, where $T$ and $V$ are their kinetic and potential energy, respectively). The turbulence consists of a divergence-free Gaussian random velocity field (with different random seed for each simulation of the set), following a \citet{Burgers48} power spectrum.  
The gas thermodynamics has been treated by adopting an adiabatic equation of state with the addition of radiative cooling, as described in \citet{Boley09} and \citet{Boley10}. The amount of energy lost by cooling was calculated through the divergence of the flux $\nabla \cdot F_{cool}=-(36\pi)^{1/3}s^{-1}\sigma(T^4-T_{irr}^4)(\Delta\tau+1/\Delta\tau)^{-1}$, where $\sigma$ is the Stefan-Boltzmann constant, $T_{irr}$ represents the irradiation temperature, $s=(m/\rho)^{1/3}$ and $\Delta\tau=sk\rho$, where $m$ and $\rho$ are the gas particle mass and density and $k$ is the local opacity. For $k$, we used Planck and Rosseland dust opacities, taken from \citet{Dalessio01}, while we chose an irradiation temperature $T_{irr}=10$ K.
Stellar feedback was not directly included in this set of simulations. 

All the simulations have a fixed number of initial SPH particle of $10^{7}$, leading to a mass resolution of $10^{-3}$ to $10^{-2} M_{\odot}$. Star formation is implemented through a sink particle algorithm adopting the same criteria as in \citet{Bate95}. In particular, this algorithm requires that i) the gas particles forming a sink have a density higher than a certain threshold, ii) $E_{\rm th}/|E_{\rm p}|<1/2$ (where $E_{\rm th}$ and $E_{\rm p}$ are the thermal and potential energy of a clump of gas particles), iii) $(E_{\rm th}+E_{\rm rot})<1$ (where $E_{\rm rot}$ is the rotational energy), iv) their total energy and the divergence of their accelerations are negative. These conditions ensure that the gas particles are actually collapsing. A further criterion requires that a sink particle is formed by at least 64 gas particles. This last criterion naturally sets a lower limit to the mass of the sinks, dependent on the adopted mass resolution of each simulation (in our case, this value ranges from $6.4\times 10^{-2}$ to $6.4 \times 10^{-1} M_{\odot}$). For our set, we chose a density threshold of $10^{7}$ cm$^{-3}$, which ensures that sink formation occurs in the highest density peaks generated by the gas collapse. After formation, sink particles can also accrete further gas particles if these end up within a sphere with radius $r_{\rm sink}=2\times 10^{-3}$ pc and satisfy the same criteria described above for the sink formation.

\subsection{From hydrodynamical to \textit{N}-body simulations: caveats}\label{caveats}

Adopting our joining/splitting algorithm to hydrodynamical sink particles, to generate initial conditions for \textit{N}-body simulations, is based on a certain number of assumptions. First of all, we assume that gas is instantaneously removed from our simulations. While there is evidence that most young star clusters are affected by gas removal, showing clear signatures of expansion \citep[see, e.g.,][]{Kuhn19,Cantat-Gaudin19,Pang20}, it is not fully clear whether this expansion is the result of an almost instantaneous gas removal by supernovae or of a continuous pre-supernova feedback. Even though we did not include pre-supernova feedback in our hydrodynamical simulations, \citet{Dale15} have shown that the pre-supernova gas removal is expected to play a minor effect on the survival and dynamics of stellar clusters (but see, e.g., \citealt{Chevance20b, Chevance20}).

This is also confirmed by the work of \citet{Parker13} and \citet{Parker15}, who performed \textit{N}-body simulations of the evolution of sink particles obtained by hydrodynamical simulations with and without pre-supernova feedback. These authors found no significant difference between these two types of simulations, showing that the instantaneous gas removal should be the main contributor to the gas-free evolution of young star clusters.

We estimated the impact of gas removal for our star clusters (whose main properties are summarized in Table \ref{tab1}), by looking at the gas fraction at the end of the hydrodynamical simulations. Figure \ref{mass} shows the gas-to-total mass fractions of our simulations, calculated for the total cluster (blue circles), within the 50\% stellar Lagrangian radius $r_{50}$ (orange triangles) and within the 10\% stellar Lagrangian radius $r_{10}$\footnote{$r_{50}$ and $r_{10}$ are defined as the radii of the spheres containing 50\% and 10\% of the mass of the gas-free star clusters, respectively. These spheres are centered on the highest density stellar peak, corresponding to the center of the main  sub-cluster.} (green squares). This plot clearly shows that even though the amount of removed gas accounts for about 60-70\% of the total initial cloud mass, this becomes 20-40\% within $r_{50}$ and <20\%  within $r_{10}$. As a further check, we calculated the kinetic and potential energy of the gas alone, for the simulation SC2, at the time of its removal. We found that its potential energy budget is about $-6\times 10^{47}$ erg, which is less than 5\% of the potential energy stored in the sink/star particles, and its total energy is positive, for a gas-only virial ratio of about 2.5 (here defined as the ratio between kinetic and potential energy of the gas). Thanks to gas cooling and self-gravity, the potential energy that is transferred to the sink particles for the duration of our hydrodynamical evolution is also about twice the initial potential energy of the cloud. As a consequence, we expect some impact of gas removal on the evolution of our star clusters, but not on their overall survival and energetic state, as already shown by \cite{Kruijssen12}, from the analysis of simulations with a setup similar to ours.

Choosing at what evolutionary time of the hydrodynamical simulation to apply our joining/splitting algorithm is also an important issue. We instantaneously remove all the gas from the simulations and we apply our splitting algorithm at $t_{\rm hydro}=3$ Myr. This choice of $t_{\rm hydro}$ was driven by the fact that at 3 Myr all the clouds converted about 30-40\% of their gas mass into stars. This is in agreement with previous results of hydrodynamical simulations including pre-supernova stellar feedback \citep[e.g.,][]{VazquezSemadeni10, Dale14, Dale15, Gavagnin17, Li19}, which show that the star formation efficiency tends to saturate to these values. This star formation efficiency and timescales are also roughly consistent with results of other works focusing on cluster survival \citep[e.g.][]{Kroupa01b,Bastian06,Goodwin06,Pfalzner14} and it is compatible with the timescale of gas dispersal and star formation quenching found by \citet{Chevance20b} and \citet{Chevance20}.


\subsection{\textit{N}-body simulations}

Figure \ref{map_stars} shows a projection of the initial spatial distributions of stars for the star clusters in our set.

We evolved these star clusters with an updated version of the direct \textit{N}-body code HiGPUs \citep{Capuzzo-Dolcetta13}, adapted to fully exploit the new AVX-512 set of instructions for recent Intel architectures. In all the simulations, we adopt a softening parameter $\epsilon=10^{-5}$ pc. 

We ran our \textit{N}-body simulations for at least 2 Myr, though for the simulations with the lowest number of particles we evolved our clusters up to 10 Myr. After 2 Myr of evolution of our set, the median value of the relative errors on energy and angular momentum conservation is $1.6\times 10^{-5}$ and $1.1\times 10^{-6}$, respectively.


\begin{figure}
\begin{center}
\includegraphics[scale=0.5]{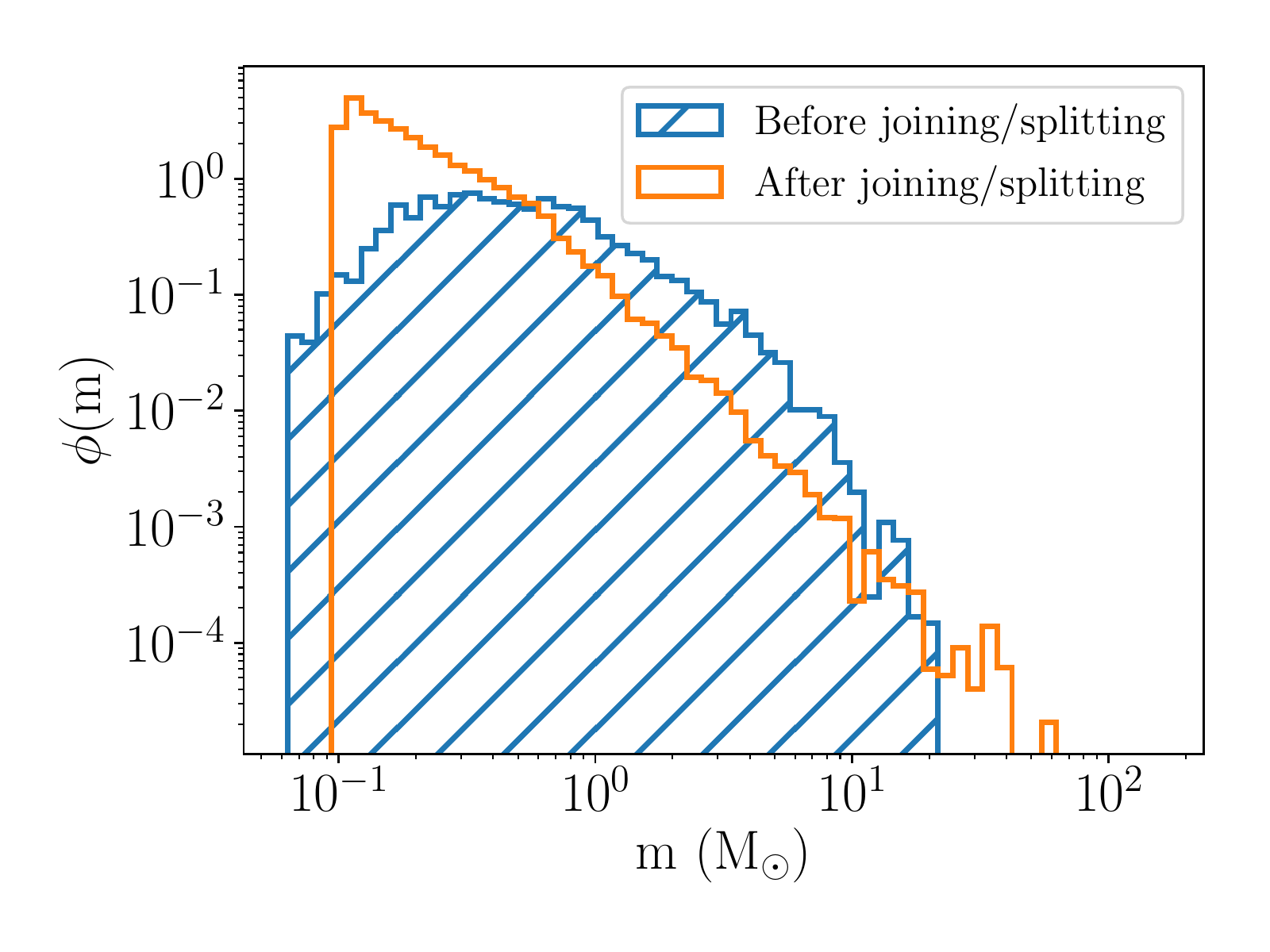}
\includegraphics[scale=0.5]{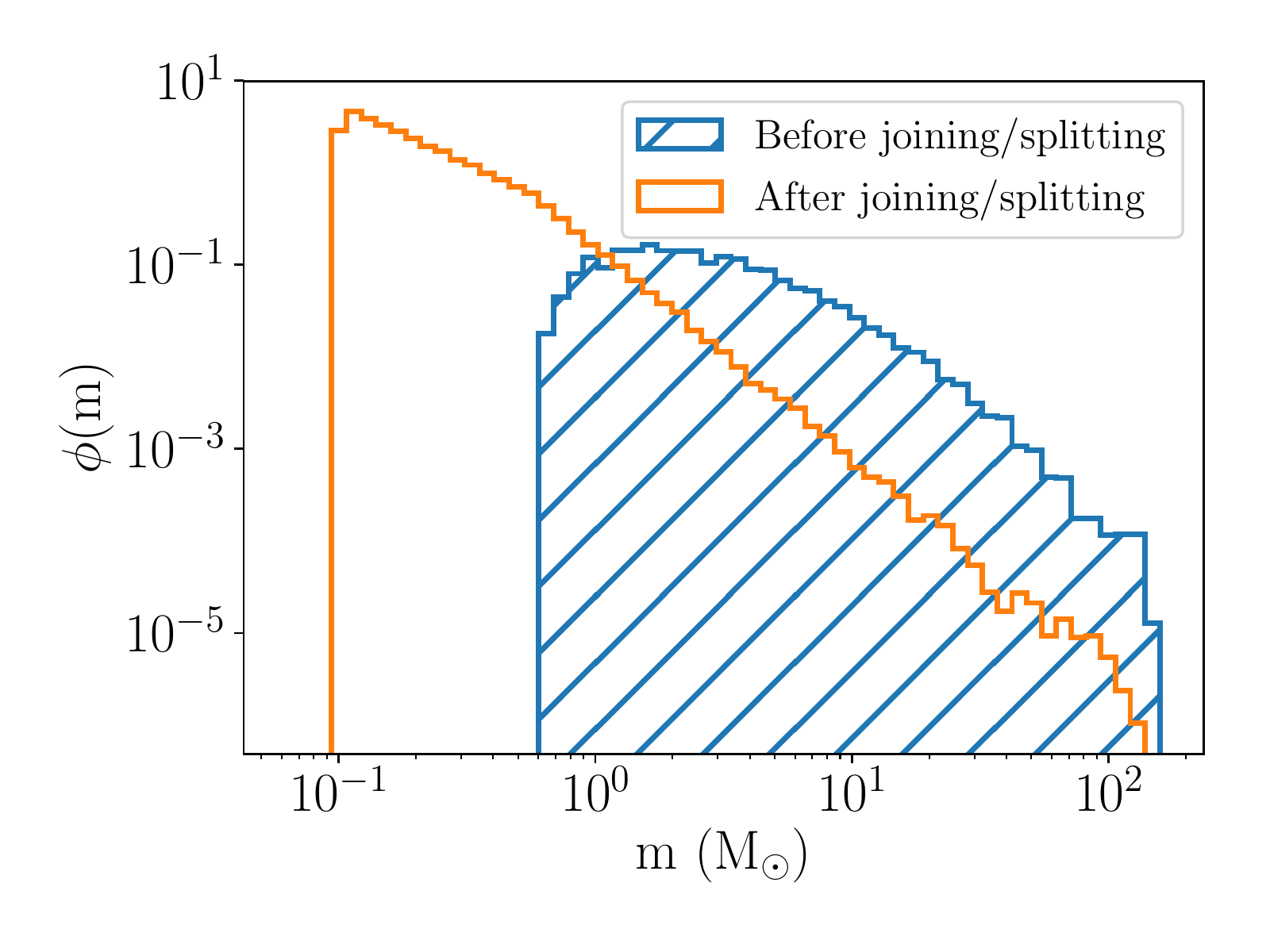}
\caption{Mass function of the original sink particles (blue histograms) and of the corresponding generated stars (orange histograms) for the SC1 (upper panel) and SC10 (lower panel) simulations.
}\label{fig_imf}
\end{center}
\end{figure}


\begin{figure*}
\begin{center}
\includegraphics[scale=0.5]{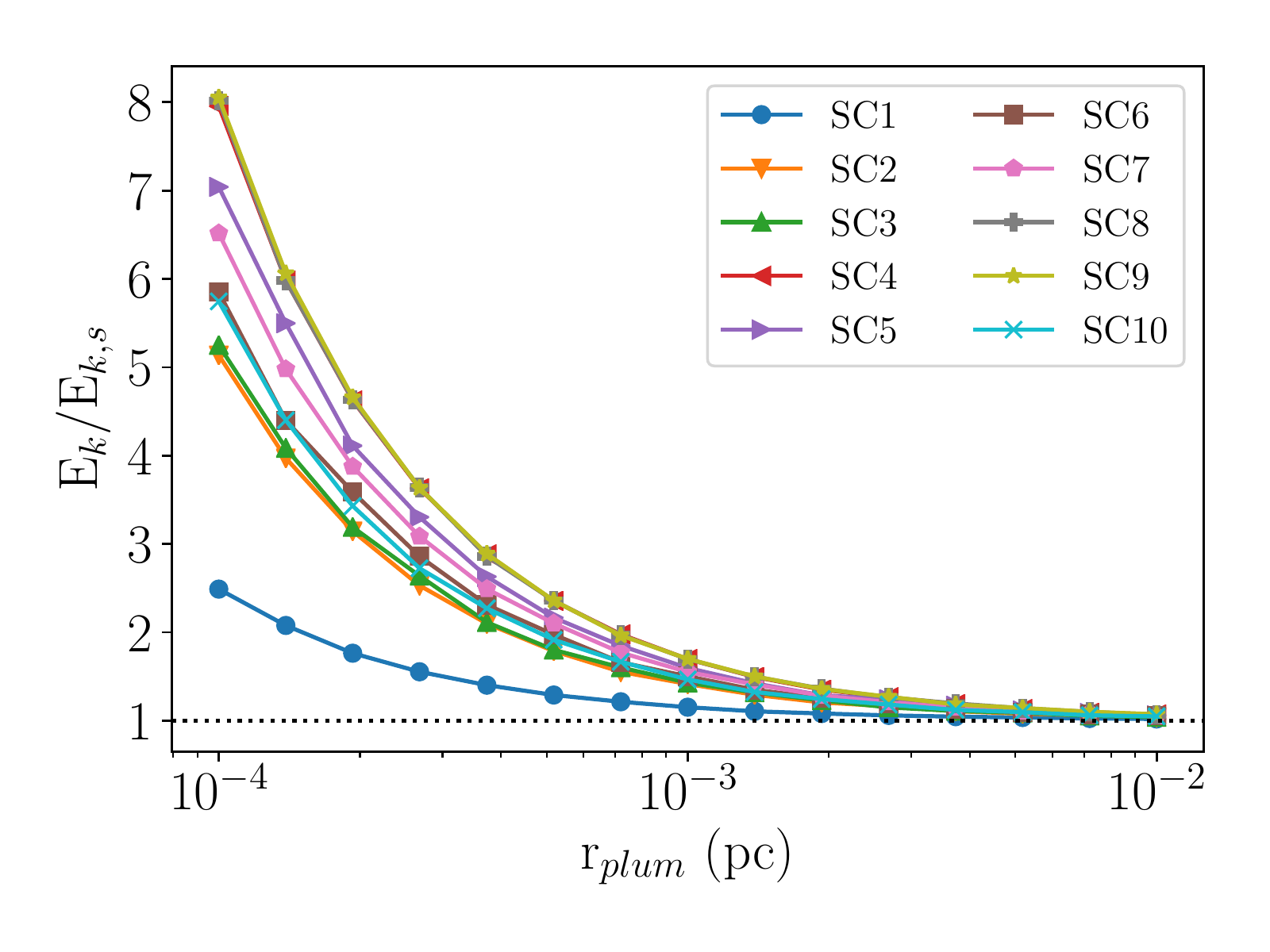}
\includegraphics[scale=0.5]{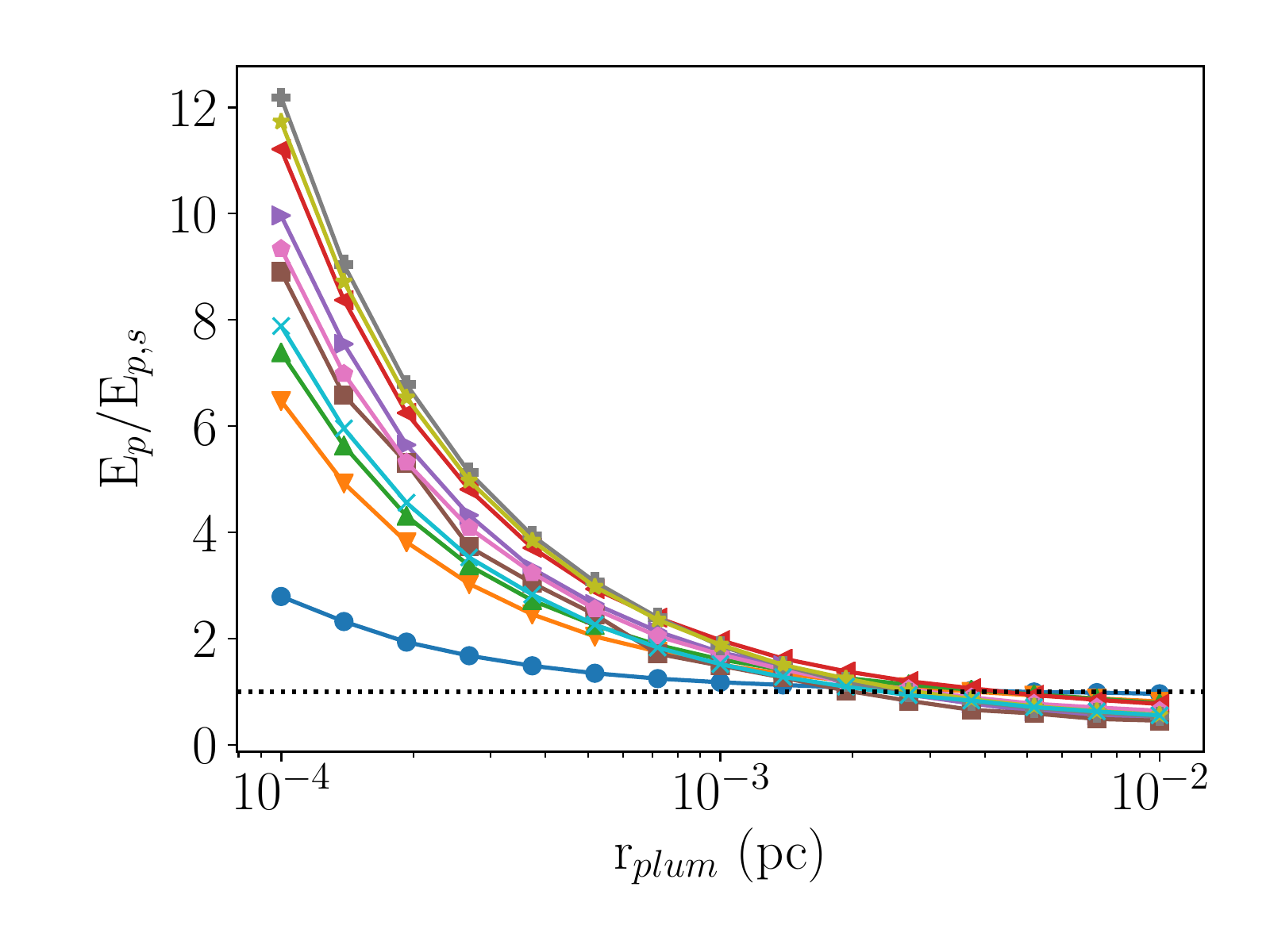}
\includegraphics[scale=0.5]{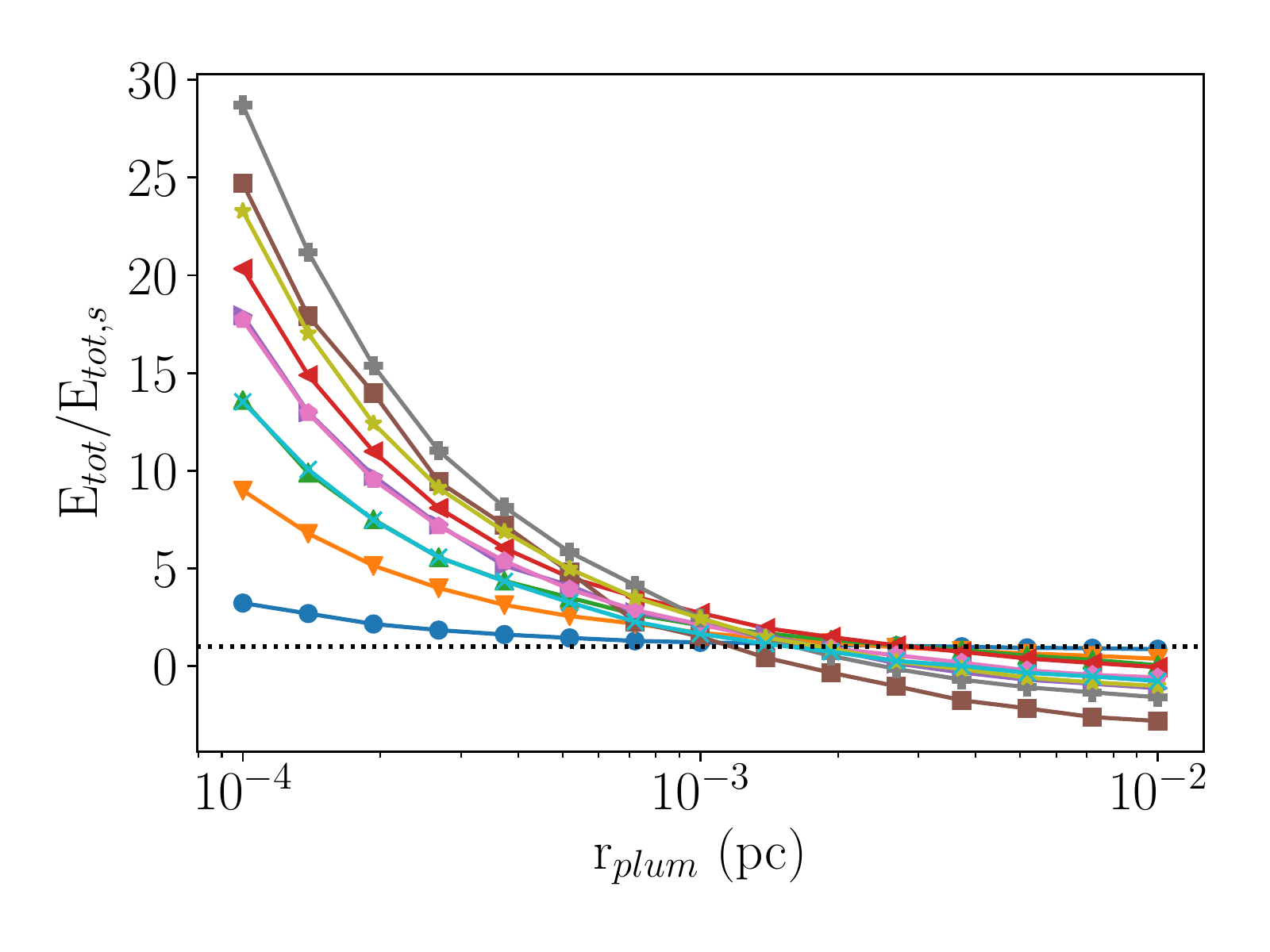}
\includegraphics[scale=0.5]{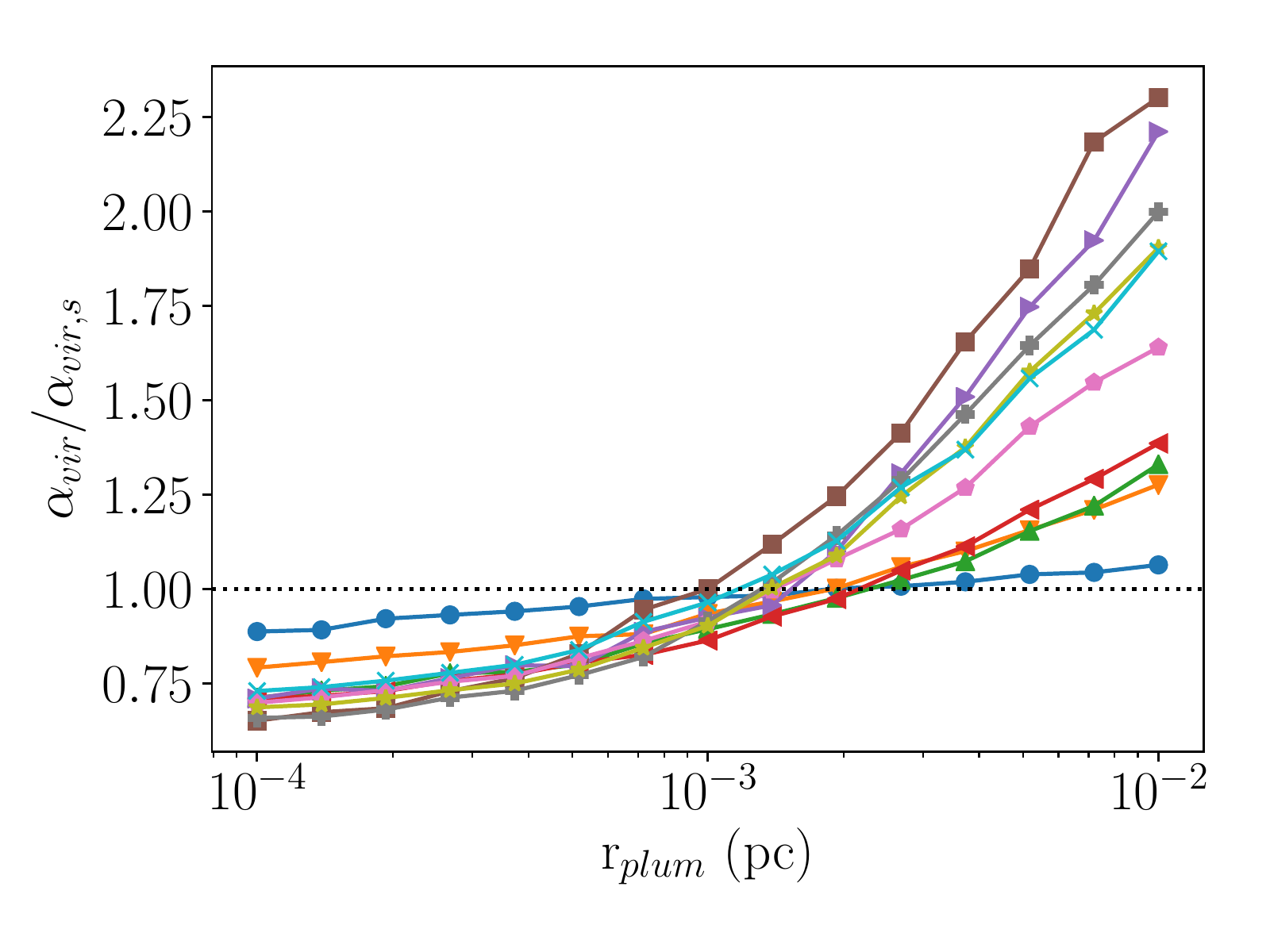}
\caption{Kinetic energy (upper left), potential energy (upper right), total energy (lower left) and virial ratio ($E_{k}/|E_{p}|$, lower right) of our set of star clusters, each scaled to the value of the original sink particles, as a function of the parameter $r_{plum}$ (see text, for details). Each point is the median of 20 different realizations of the same star cluster, after applying the joining/splitting algorithm. The black dotted horizontal line represent the original sink reference values.
}\label{var_rplum}
\end{center}
\end{figure*}

\section{Tests}\label{tests}

In this section, we discuss the performance of our algorithm, mainly by means of a comparison between the mass function, energy conservation and particle separation before and after the joining/splitting.

\subsection{Mass function}\label{mf_sec}

Figure~\ref{fig_imf} shows the mass function of the original hydrodynamical sink particles and of the generated stars, for the SC1 and SC10 star clusters. For all the following results, we adopted a Kroupa mass function \citep{Kroupa01}, with lower and upper end of 0.1 and 150 M$_{\odot}$, respectively. We set the lower limit to 0.1 M$_\odot$, since including smaller masses would have dramatically increased the computational cost of the simulations and would have had a small impact on the dynamical evolution of the star cluster. The mass function of sink particles for the most massive case is overall shifted to larger masses. This is an effect of the different mass resolutions adopted for the set of hydrodynamical simulations described in section \ref{hydro} and in \citet{Ballone20}. 
In terms of our algorithm, this means that the stars from the SC1 cluster were generated through both the joining and splitting branches of our algorithm, while those of the SC10 cluster were born only from splitting of their parent sink particles.


\begin{figure*}
\begin{center}
\includegraphics[scale=0.5]{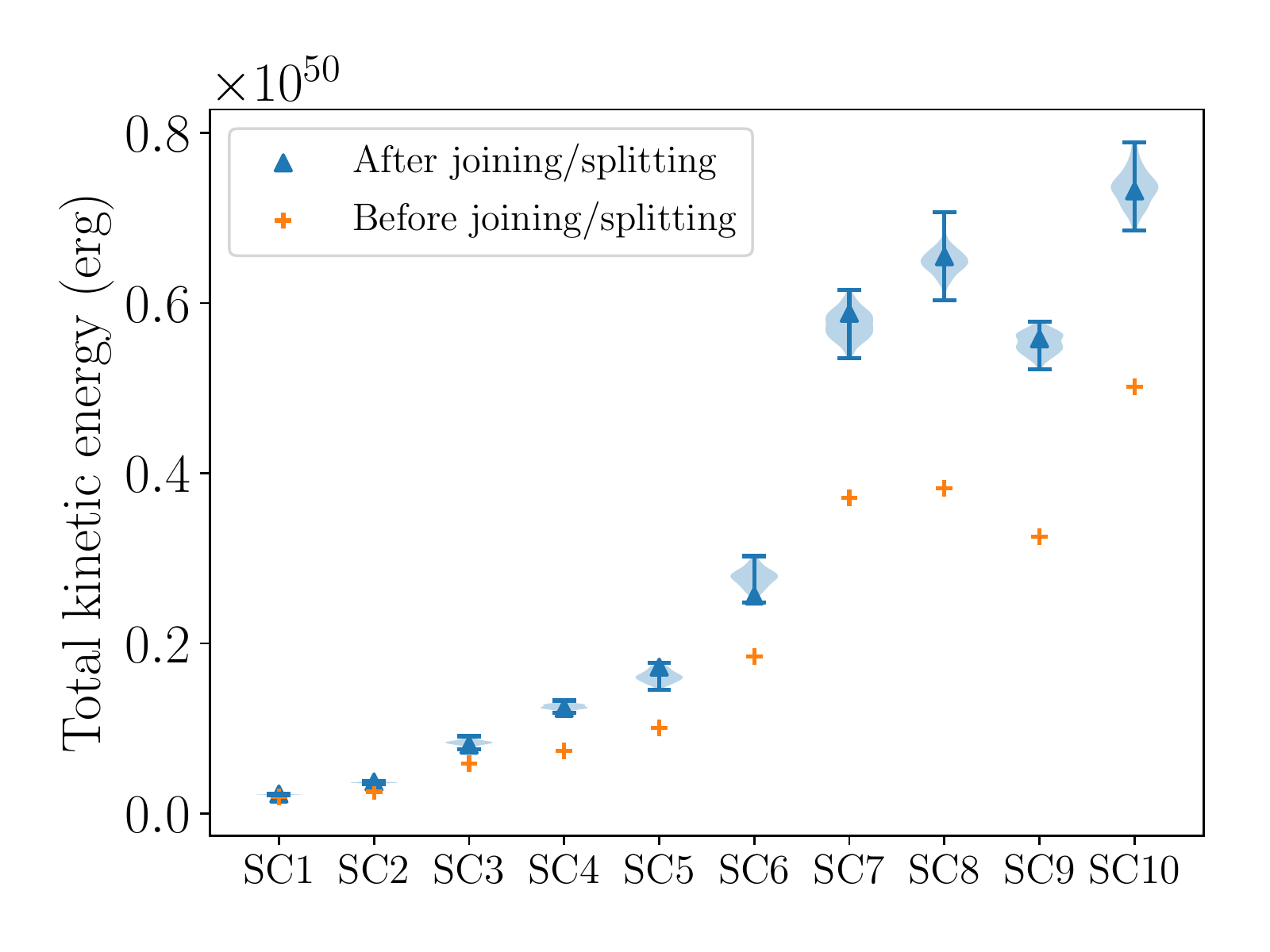}
\includegraphics[scale=0.5]{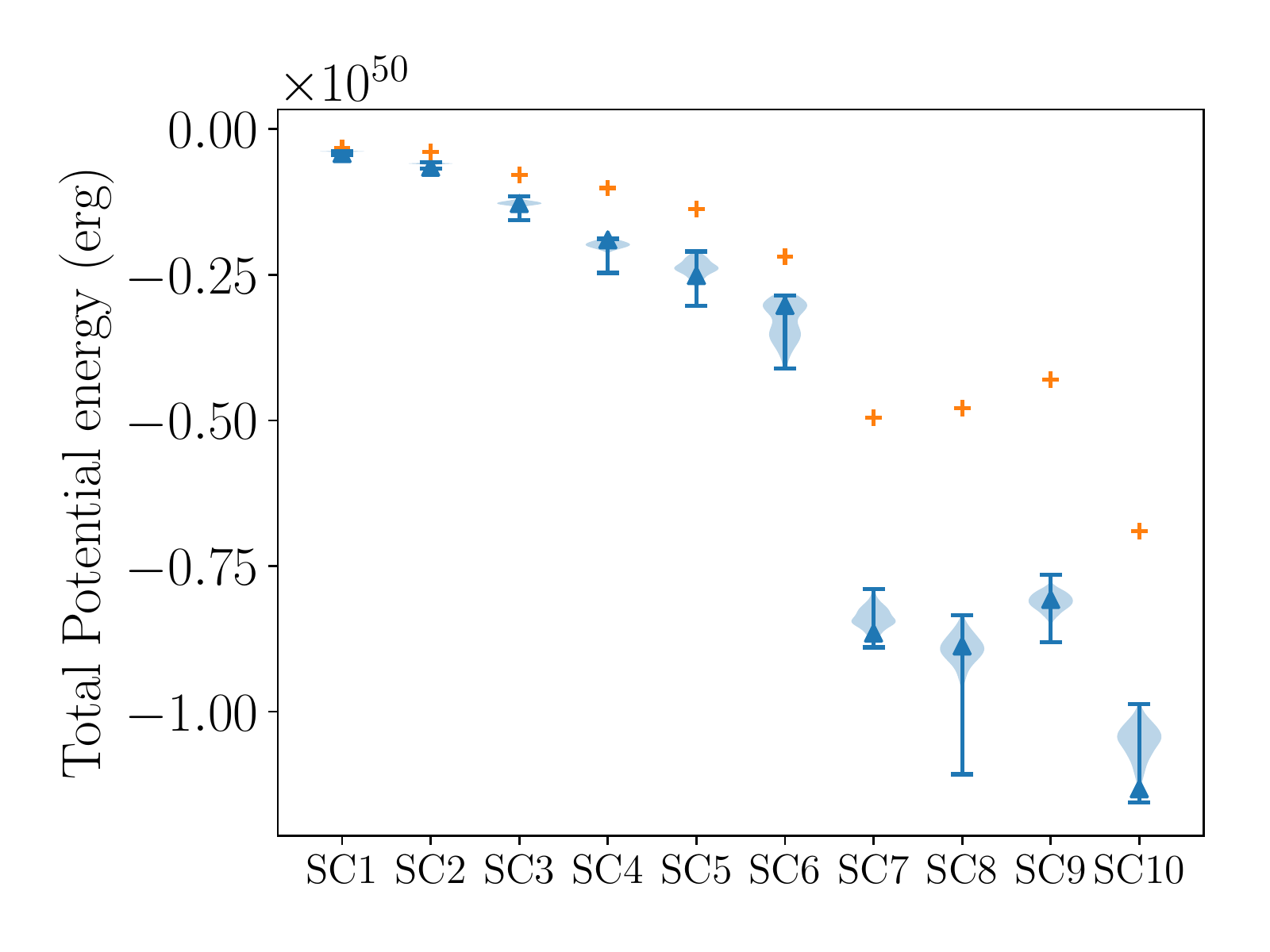}
\includegraphics[scale=0.5]{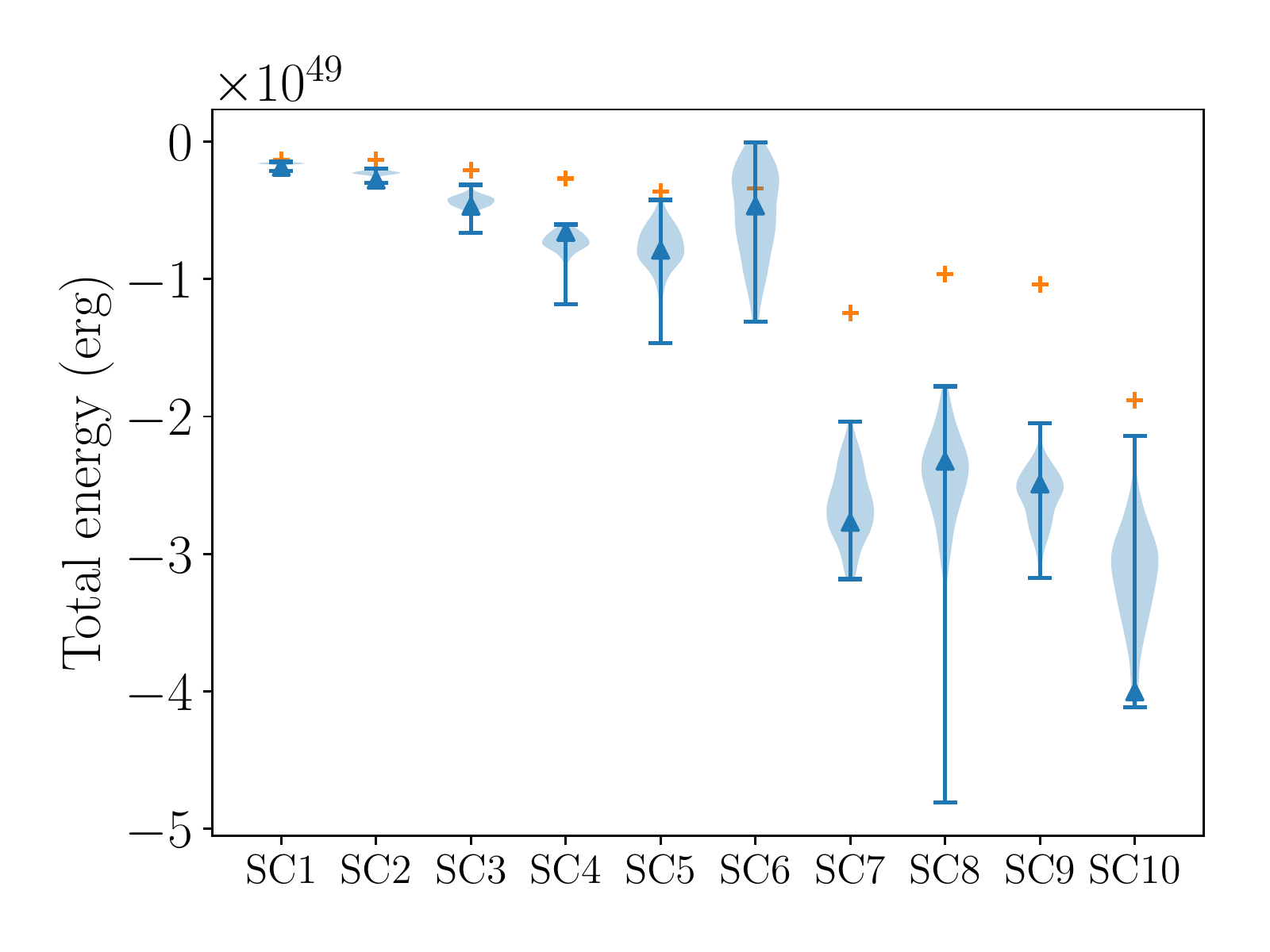}
\includegraphics[scale=0.5]{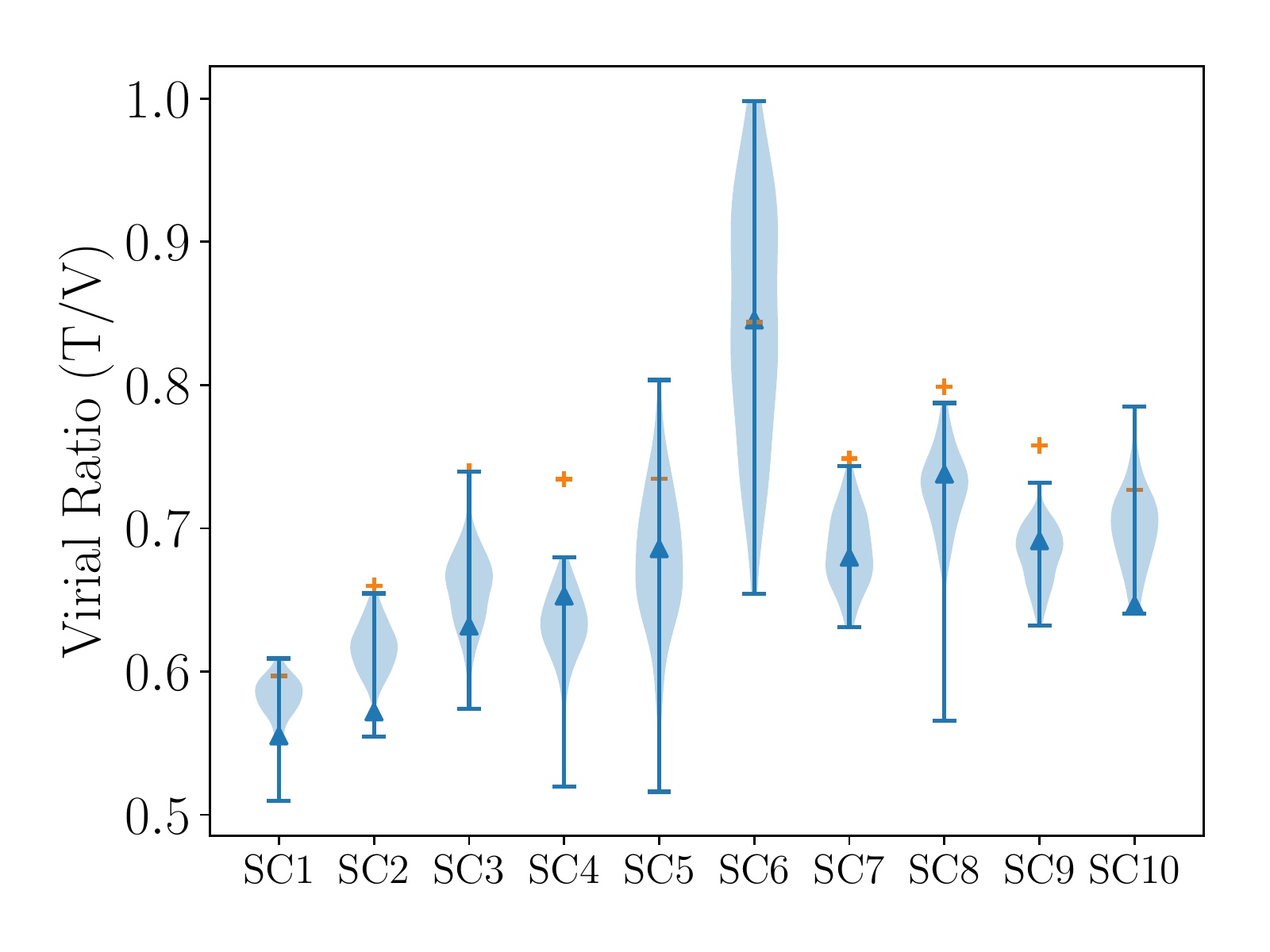}
\caption{Kinetic energy (upper left), potential energy (upper right), total energy (lower left) and virial ratio ($E_{k}/|E_{p}|$, lower right) for 100 realizations of our joining/splitting applied to each hydrodynamical simulation. The blue triangles represents the values for the initial conditions adopted for our set of \textit{N}-body simulations (see Table \ref{tab1}), the violin plots show the distributions of the 100 different realizations. The orange crosses show the values for the sink particles in the hydrodynamical simulations (before applying the joining/splitting).
}\label{var_split}
\end{center}
\end{figure*}

\begin{figure*}
\begin{center}
\includegraphics[scale=0.5]{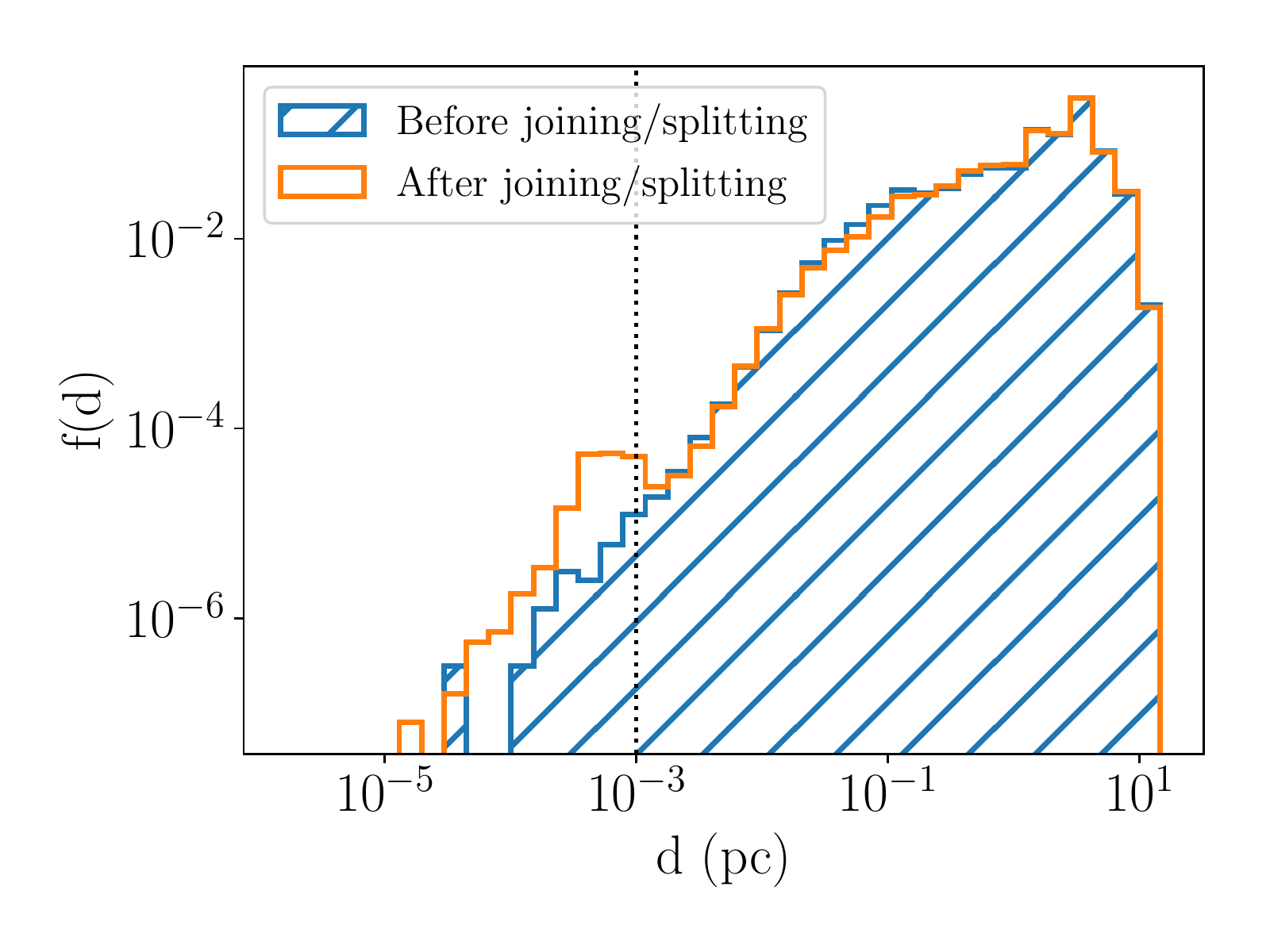}
\includegraphics[scale=0.5]{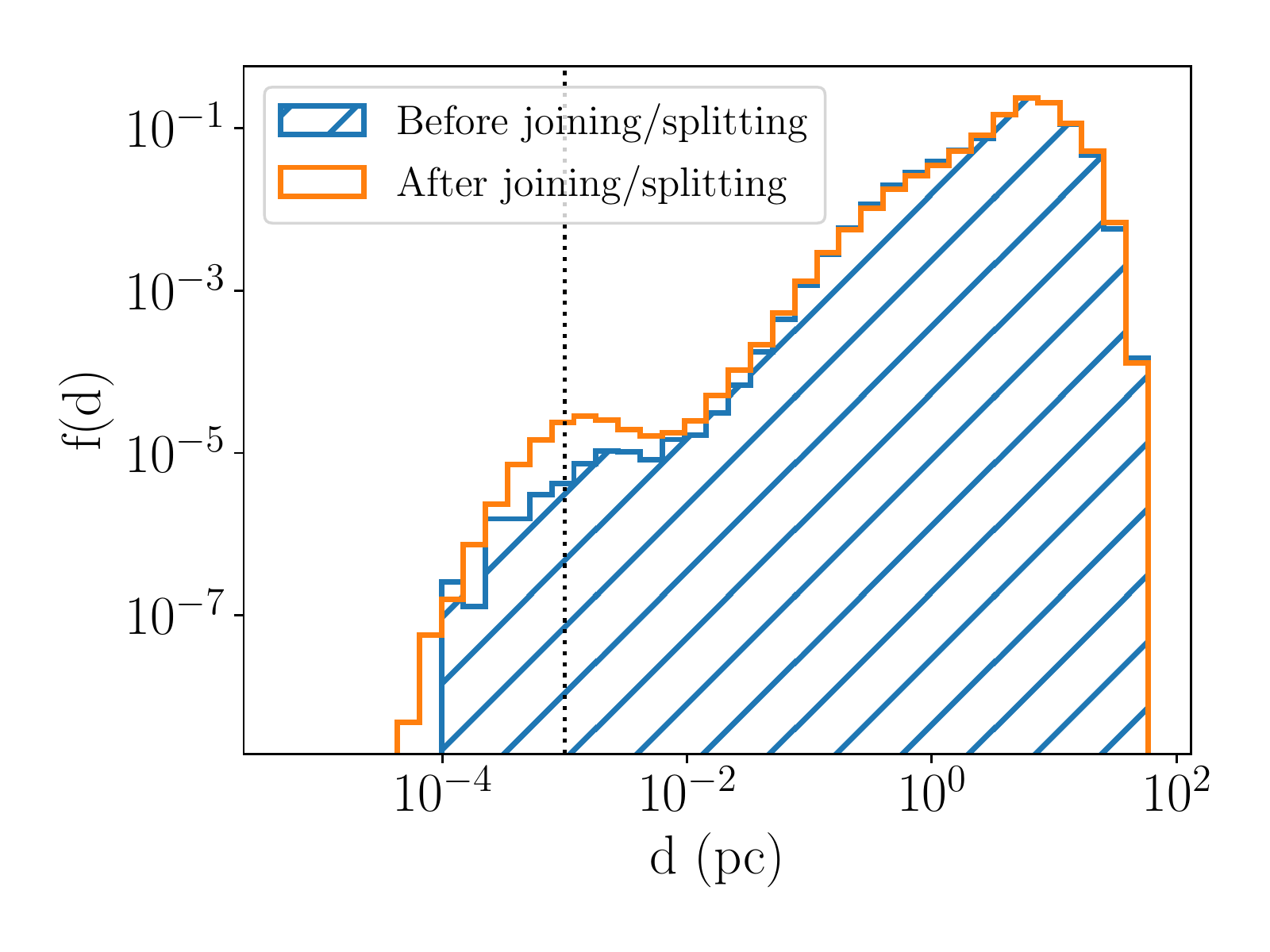}
\caption{Distribution of the particle separations for the original sink particles (blue histograms) and for the corresponding generated stars (orange histograms) for the SC1 (left-hand panel) and SC10 (right-hand panel) simulations. The vertical dotted line shows the value of $r_{plum}=10^{-3}$ pc, adopted for this joining/splitting.
}\label{fig_dist}
\end{center}
\end{figure*}


\subsection{Plummer radius of the splitting}\label{rplum_sec}

As discussed in section \ref{methods}, one of the free parameters of our algorithm is $r_{plum}$, i.e., the scale radius of the Plummer sphere over which children stars are distributed in the splitting branch. In Fig. \ref{var_rplum} we show how kinetic, potential, total energy and virial ratio vary for different choices of $r_{plum}$. As visible in the upper right-hand panel, the potential energy $E_p$ of the new stars is 
higher than that of sink particles\footnote{From now on, each quantity, when calculated for the original sink particles, will be labeled with the \textit{s} subscript.} for small values of $r_{plum}$, since it is dominated by the single Plummer spheres of the splitting. On the other hand, for large values of $r_{plum}$ all the Plummer spheres of the splitting are looser and the potential energy of the new star cluster tends to that of the original sink particles. The same argument applies to the kinetic energy $E_k$ (upper left-hand panel) and to the total energy $E_{tot}$ (lower left-hand panel): the energy of the injected Plummer spheres is the dominant term for the smallest values of $r_{plum}$, considering that each of the Plummer models is put in virial equilibrium. 

The dependence of the virial ratio (here defined as $\alpha_{vir}=E_{k}/|E_p|$) is also understandable by looking at the other curves. When $r_{plum}$ is large $E_k/E_{k,s}$ tends to 1 and $E_p/E_{p,s}$ becomes increasingly smaller than 1, hence the virial ratio becomes higher than the original virial ratio of the sinks. On the other hand, for small values of $r_{plum}$, the virial ratio of the stars tends to the virial ratio of the Plummer spheres (equal to 0.5) and $\alpha_{vir}/\alpha_{vir,s}\approx 0.5/\alpha_{vir,s}$, where $\alpha_{vir,s}$ is the virial ratio of the hydrodynamical sinks, before the splitting.

After understanding the curves in Fig. \ref{var_rplum}, it is quite trivial to realize that total energy and virial ratio cannot be simultaneously "conserved" with our joining/splitting algorithm. However, by looking at Fig. \ref{var_rplum}, it seems that for our simulations the values of these two quantities are almost conserved for $r_{plum}$ of the order of $10^{-3}$ pc, hence we adopted this value for the forthcoming analysis. Such value is of the order of some typical interparticle distance at small scales, probably driven by our choice of the sink particle radius (see Sec. \ref{hydro}).

\subsection{Randomness of the algorithm}

In order to test how the randomness of the algorithm affects the properties of the star clusters, we performed 100 different realizations of the joining/splitting procedure for each hydro-simulation and then estimated the distribution of the kinetic, potential, total energy and virial ratio.
As visible in Fig.~\ref{var_split}, the choice of $r_{plum}=10^{-3}$ pc works better for the least massive clusters (see section \ref{rplum_sec}). However, the difference in terms of energy, before and after splitting, is also partially explained by the higher difference between the mass function of the hydrodynamical sink particles and the adopted stellar mass function. In fact, for the highest cloud mass simulations the resolution of the gas particles in the hydrodynamical simulations was lower, leading to typically more massive sinks (see section \ref{mf_sec}). In this case, the splitting branch of our algorithm is dominating over the joining branch. As a consequence (as also discussed in section \ref{rplum_sec}), higher departures from the original energy and virial ratios are to be expected for these most massive simulations.
The effect of the splitting branch of the algorithm is also visible on the distributions of the kinetic, potential and total energies, which are broader for the most massive simulations.

Moreover, we see that random realizations of the same star cluster can lead to large fluctuations of the virial ratio. However, even if some extremely different realizations can be obtained, in terms of their equilibrium state, most of the realizations are distributed within less than 20\% of the mode value.

\subsection{Particle separation}\label{ref_bin}

Figure~\ref{fig_dist} shows how the interparticle distance distribution is affected by our joining/splitting. As visible, our algorithm mostly introduces new particles at distances of the order of the parameter $r_{plum}$. Nonetheless, it is interesting to notice that particles at those relative distances are already present in the simulation \citep[which is of the order of the sink radius adopted in the hydrodynamical simulation; see Section \ref{hydro} and][]{Ballone20}. Hence, for appropriate choices of $r_{plum}$, our algorithm alters only mildly the distribution of particles in space, again explaining why the potential energy of the system can be roughly preserved (see Figures \ref{var_rplum} and \ref{var_split}).

\begin{figure}
\begin{center}
\includegraphics[scale=0.5]{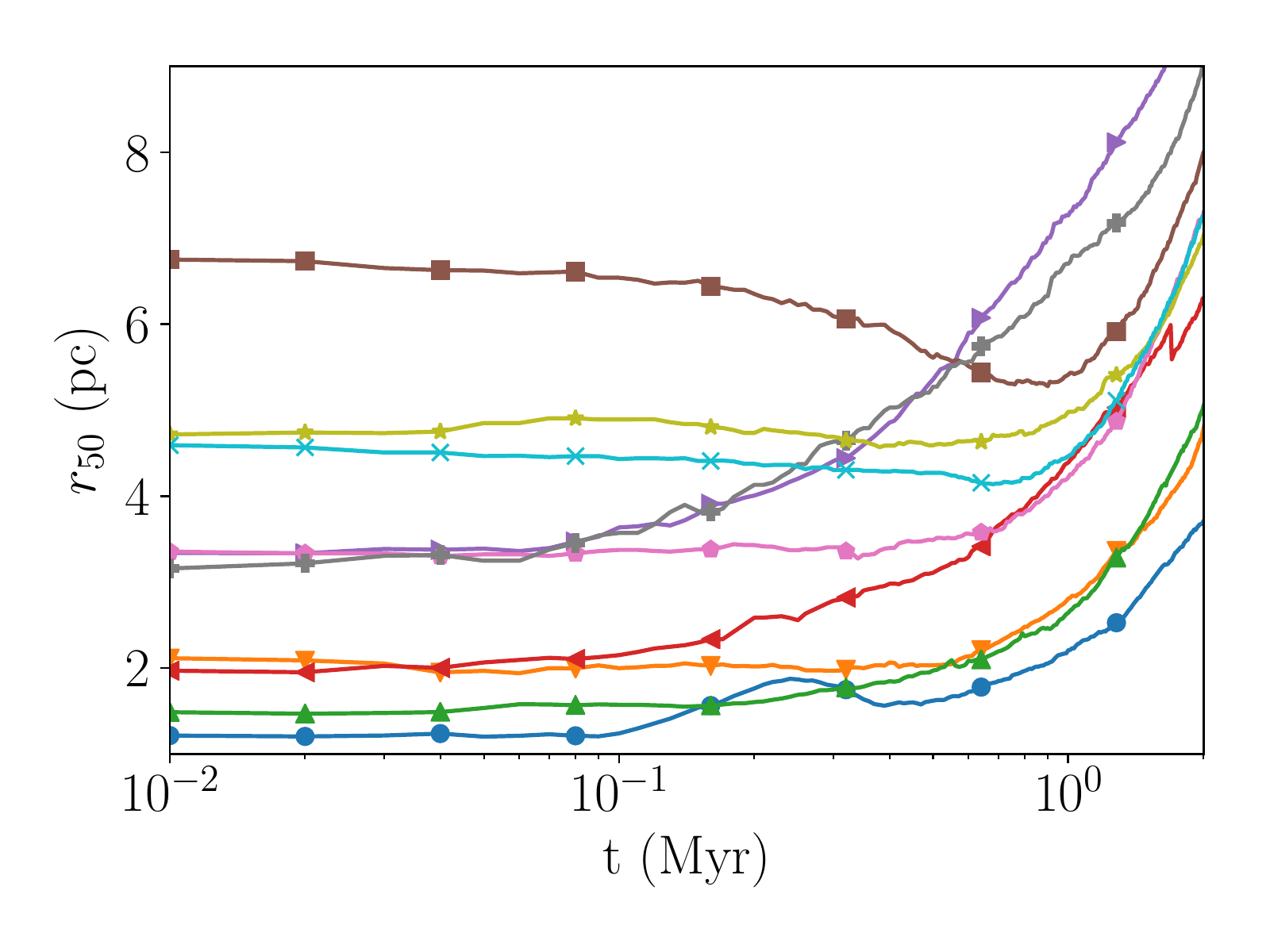}
\includegraphics[scale=0.5]{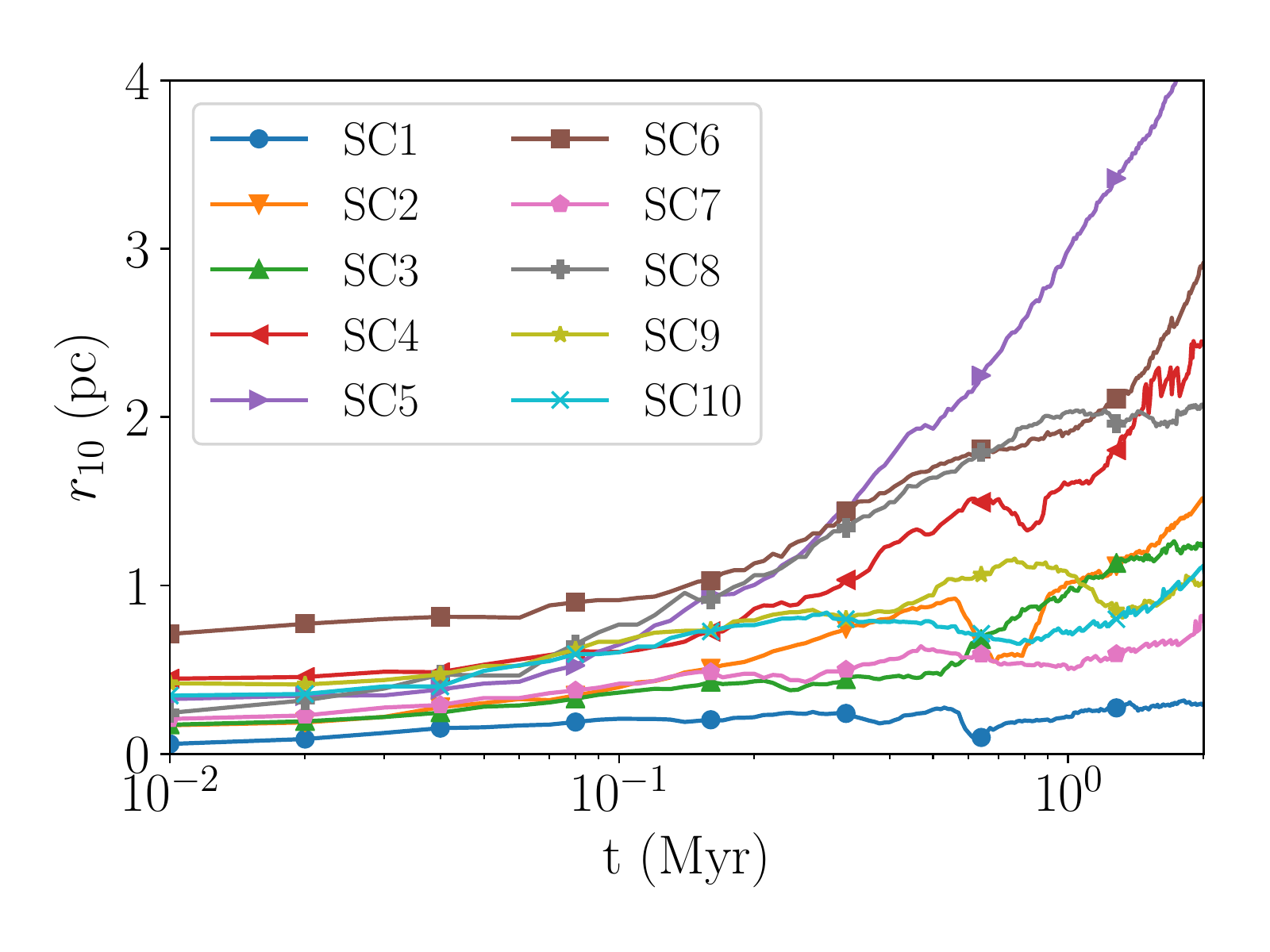}
\caption{Early evolution of the 50\% Lagrangian radius (upper panel) and 10\% Lagrangian radius (lower panel) for our set of \textit{N}-body simulations.
}
\label{radii_set}
\end{center}
\end{figure}

\begin{figure}
\begin{center}
\includegraphics[scale=0.5]{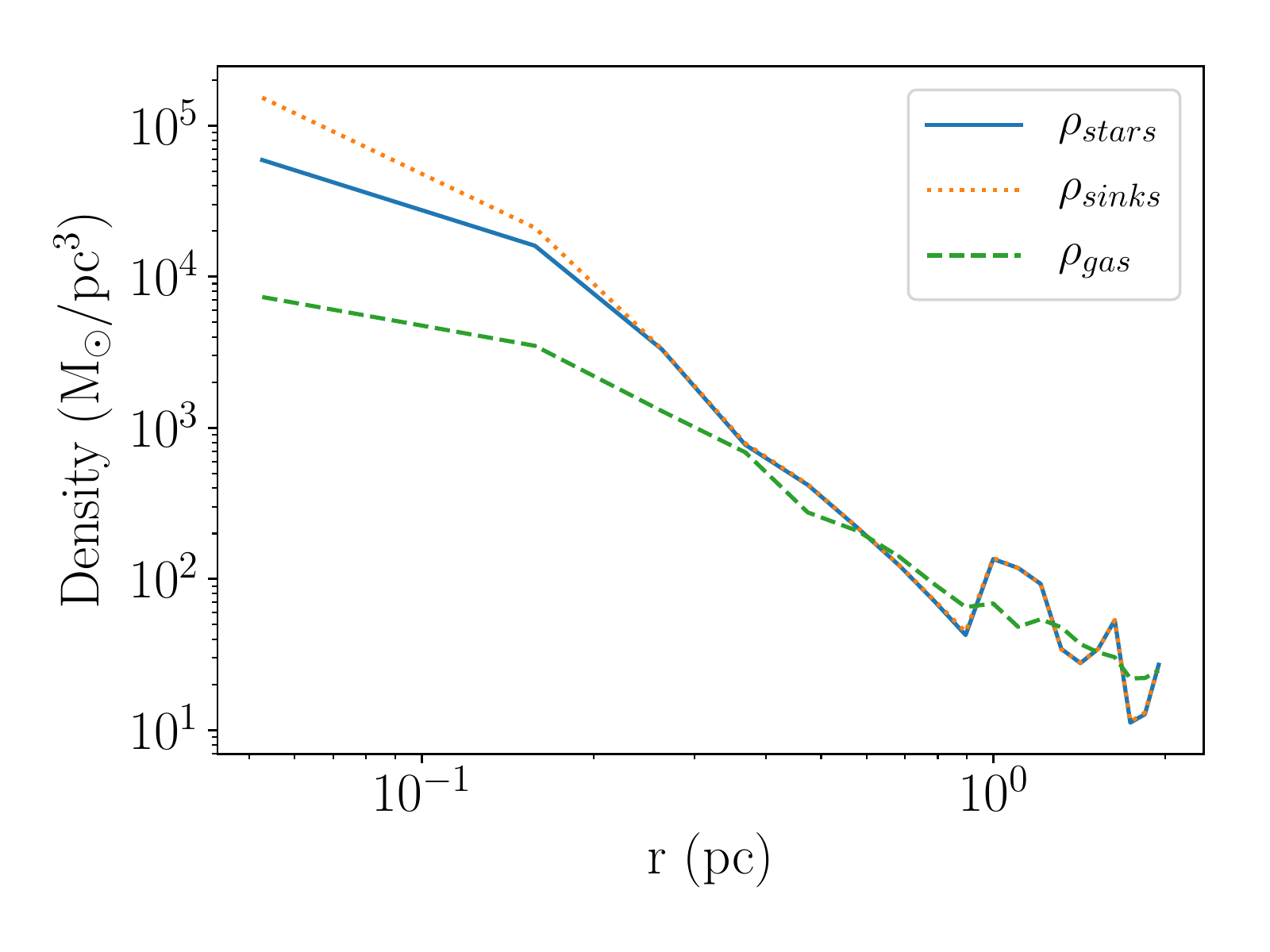}
\caption{Initial density profiles for the densest sub-cluster in the SC2 simulation. The orange-dotted and green-dashed lines show the mass density of sink and gas particles in the hydrodynamical simulation. The blue-solid line shows the corresponding distribution of stars, after the joining/splitting. 
}\label{density}
\end{center}
\end{figure}


\begin{figure*}
\begin{center}
\includegraphics[scale=0.35]{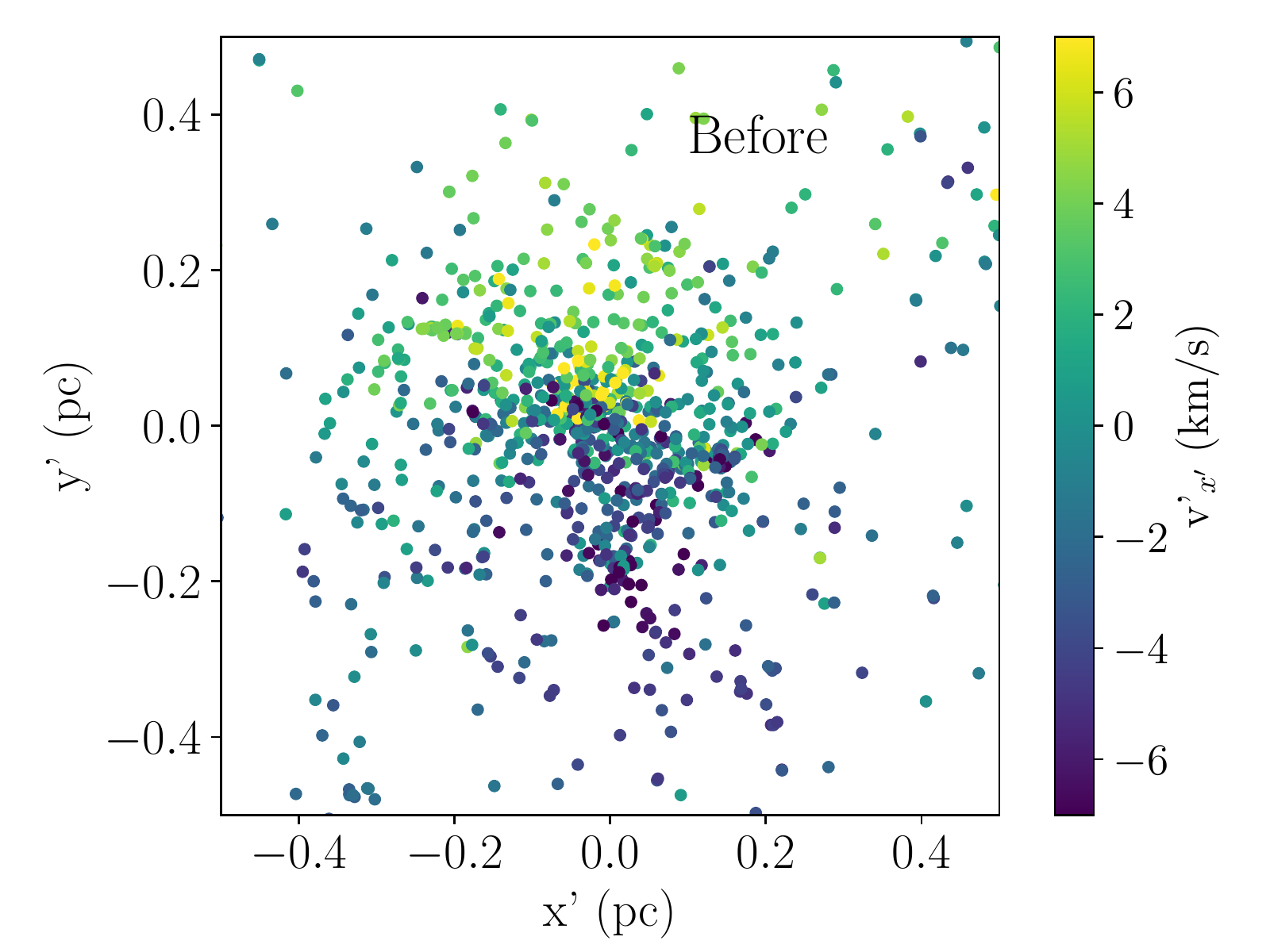}
\includegraphics[scale=0.35]{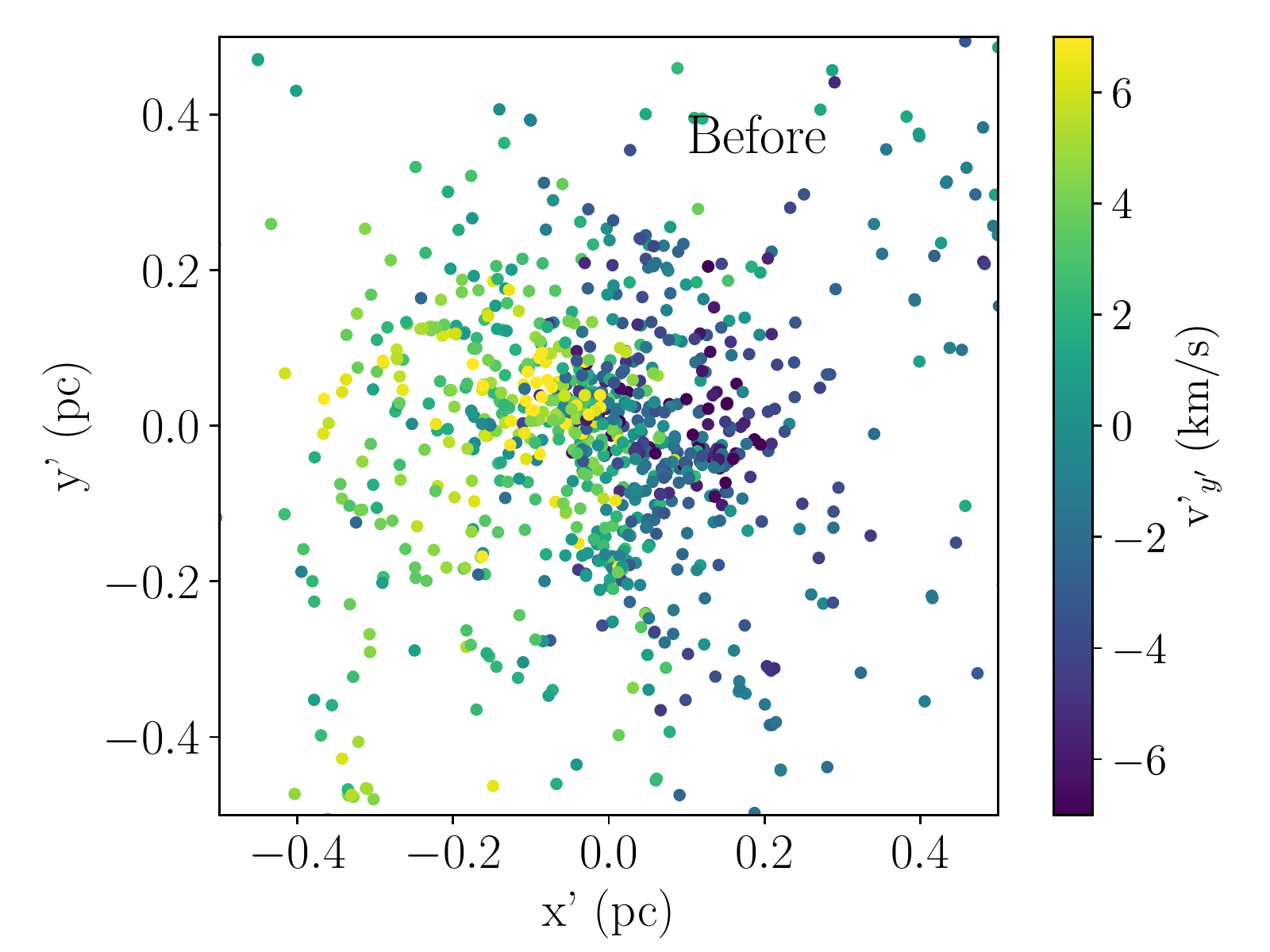}
\includegraphics[scale=0.35]{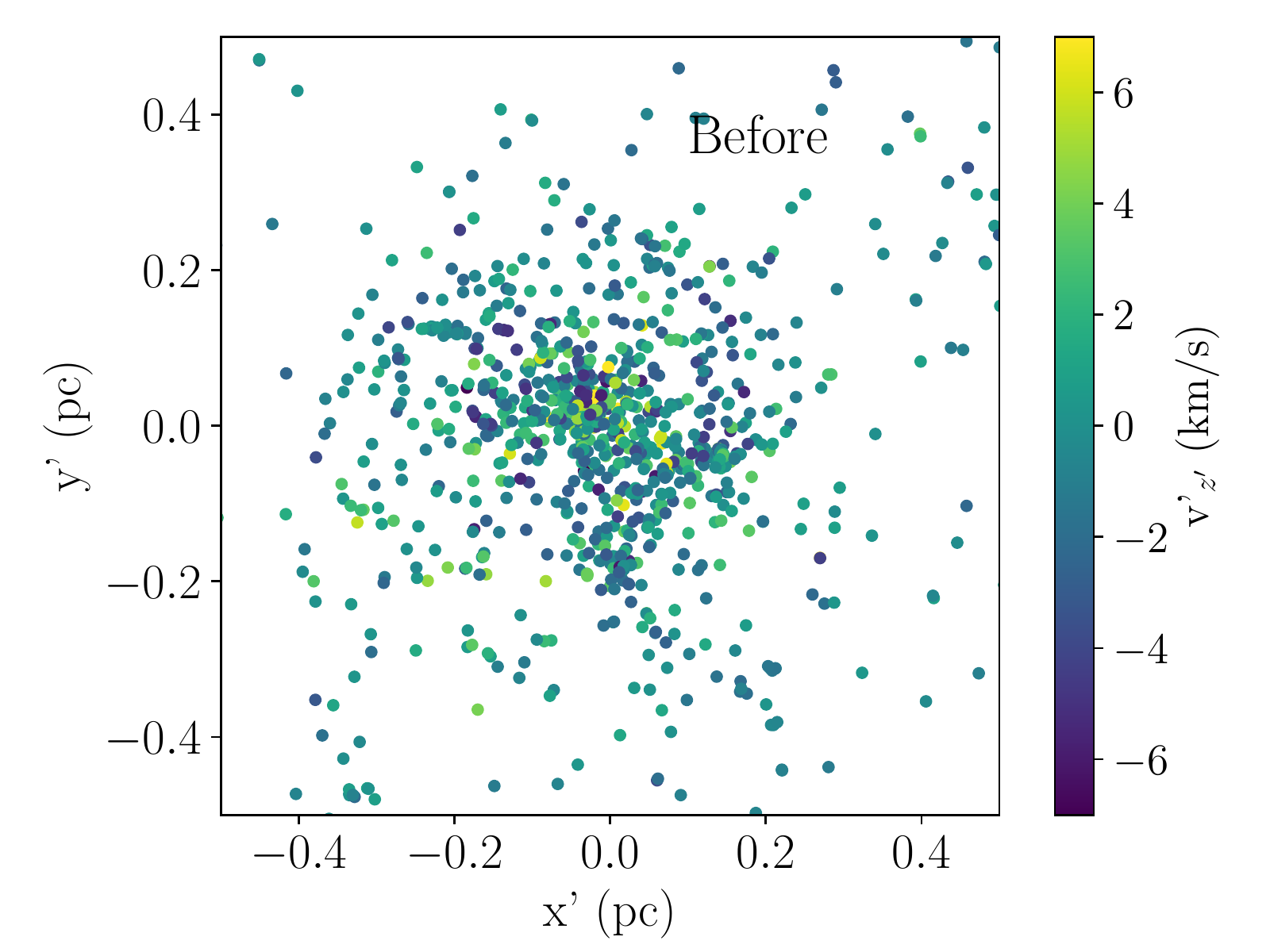}
\includegraphics[scale=0.35]{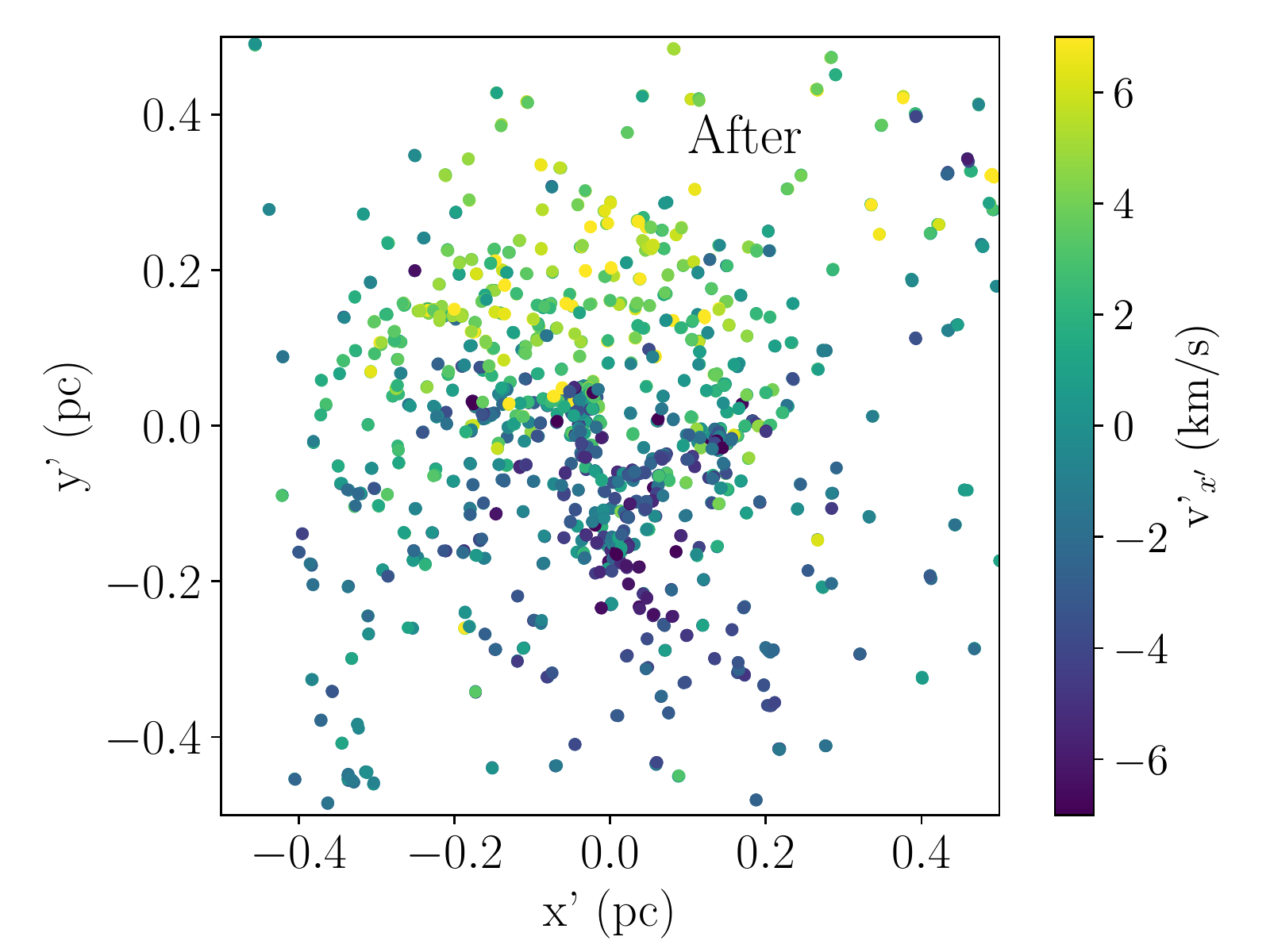}
\includegraphics[scale=0.35]{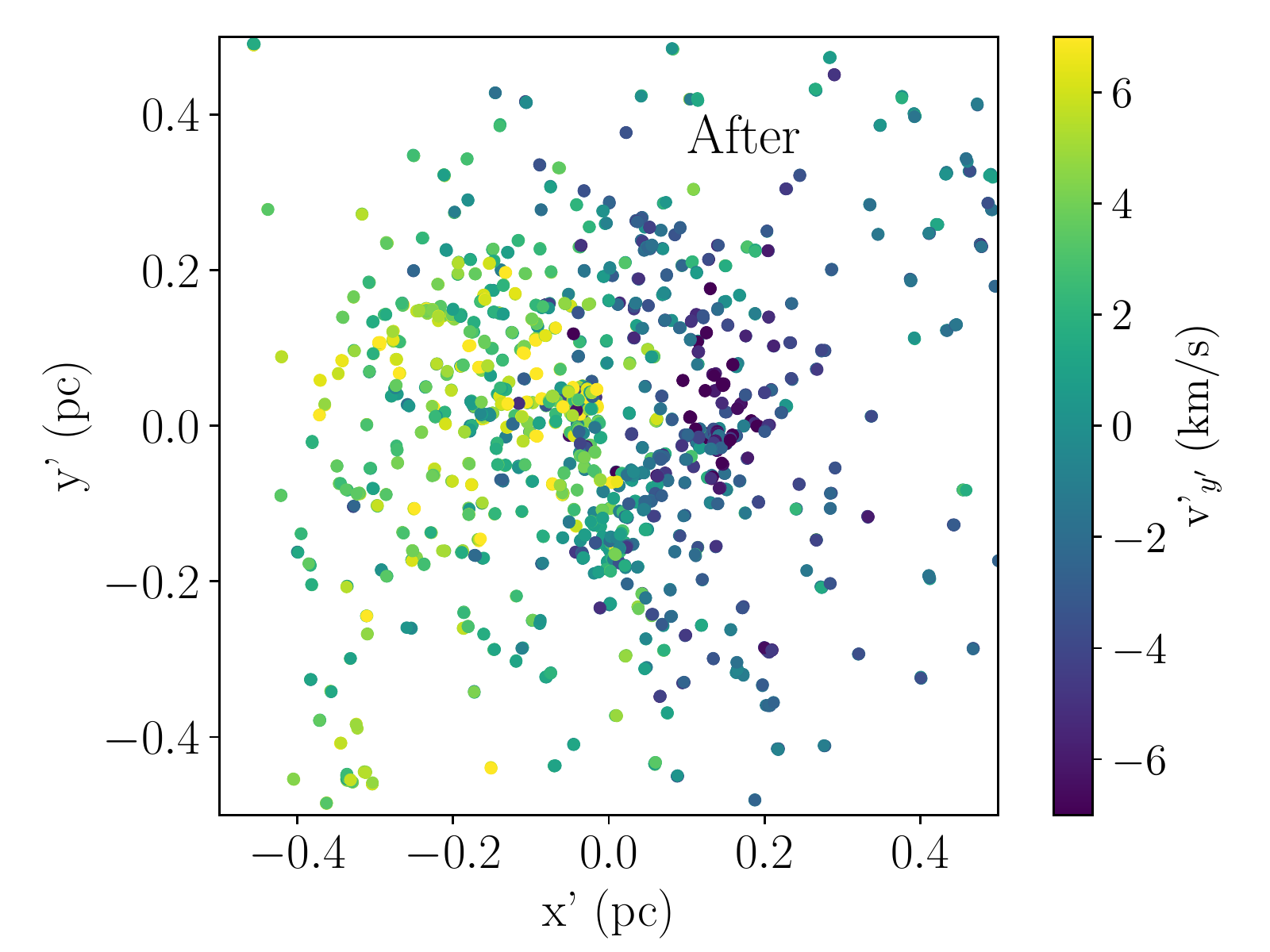}
\includegraphics[scale=0.35]{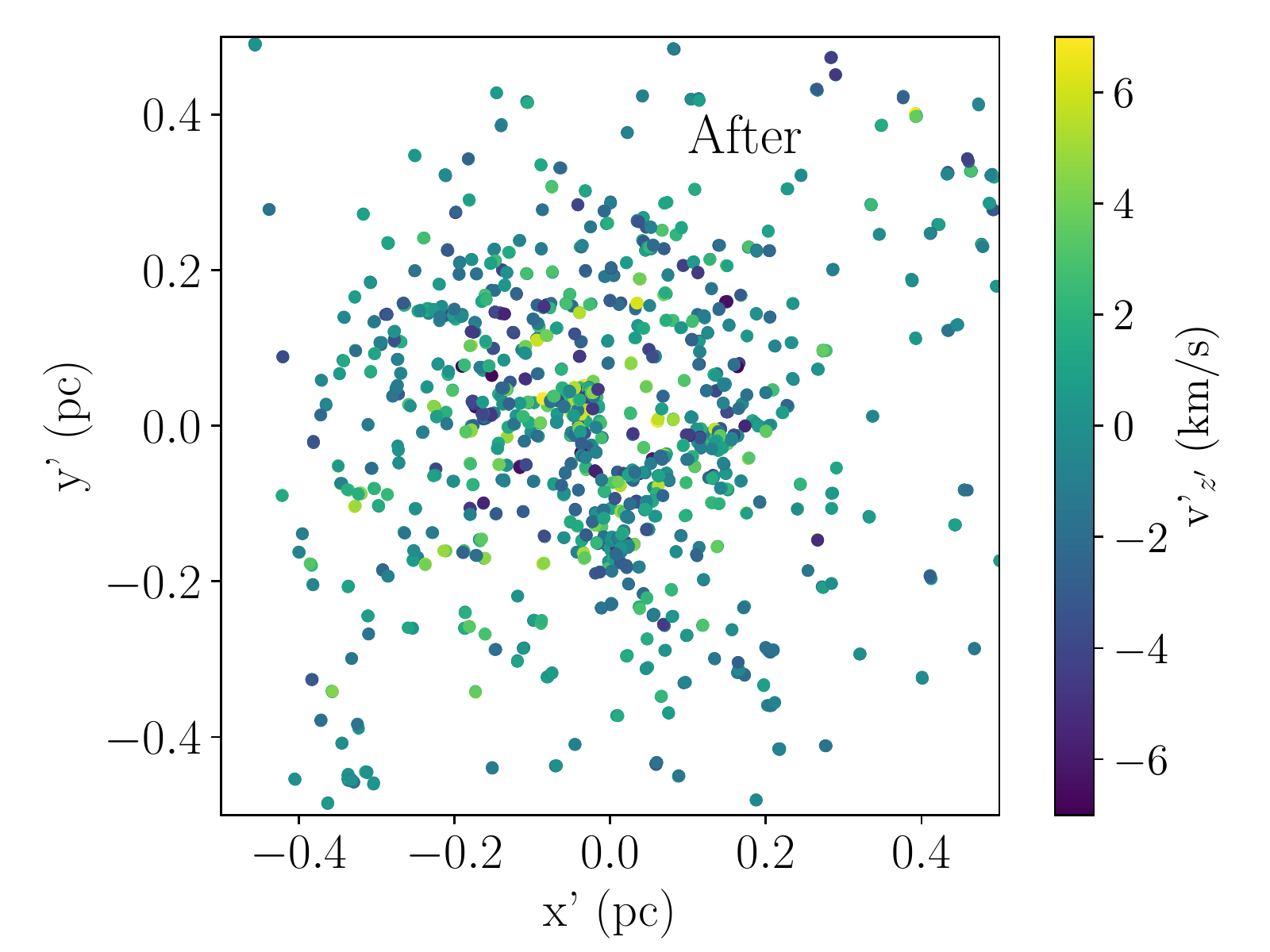}
\caption{Scatter plots of the highest angular momentum (and most massive) sub-cluster formed in the SC2 star cluster, before (upper panels) and after (lower panels) applying our joining/splitting algorithm. We show the $x'-y'$ plane, where $z'$ is the direction of the angular momentum of the sub-cluster. The colour map refers to the three components of the velocity: $v'_{x'}$ (left), $v'_{y'}$ (center) and $v'_{z'}$ (right). 
}\label{rot_comp}
\end{center}
\end{figure*}

\begin{figure}
\begin{center}
\includegraphics[scale=0.5]{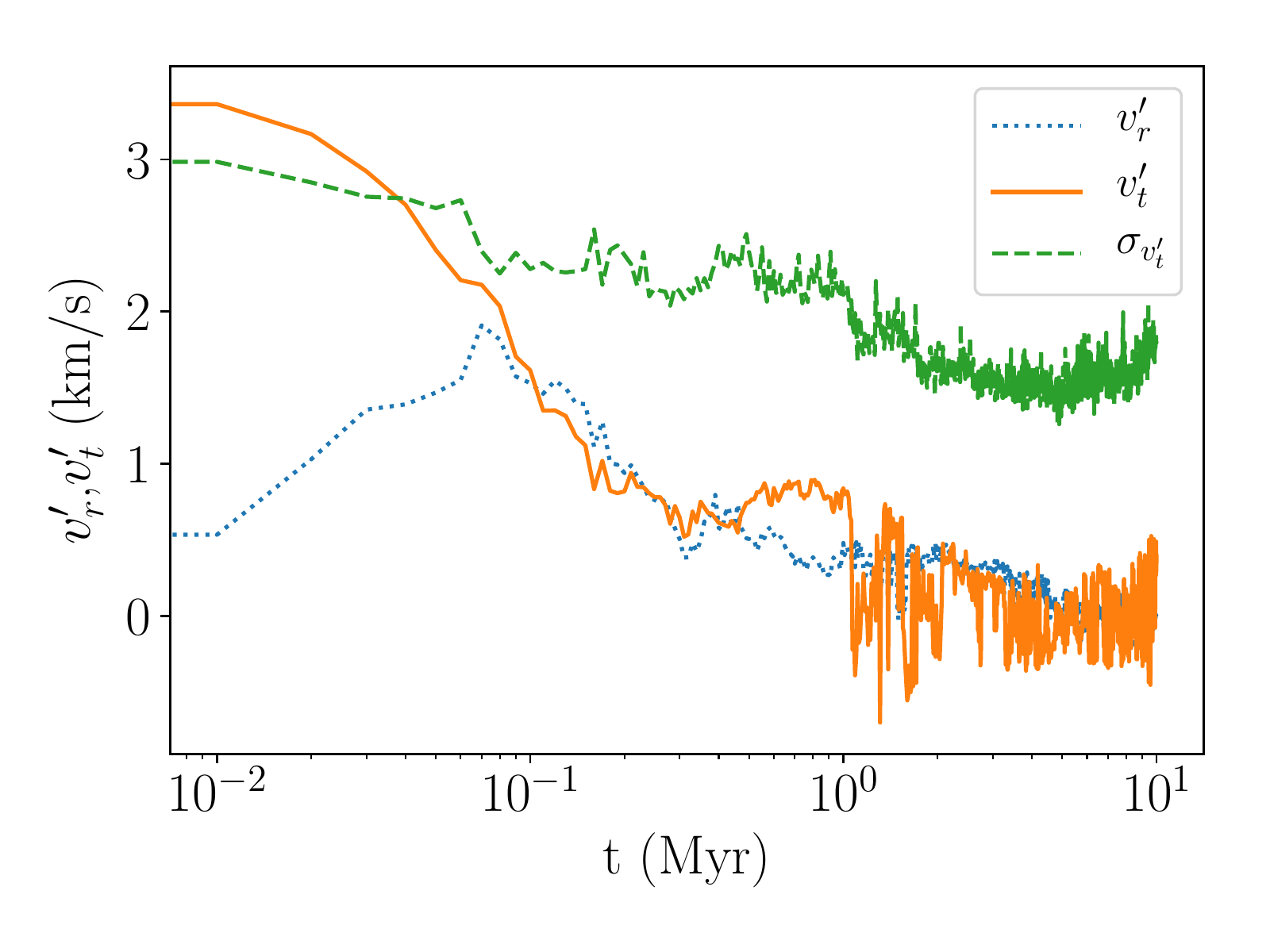}
\caption{Evolution of the tangential velocity (orange solid line), radial velocity (blue dotted line) and tangential velocity dispersion (green dashed line) for the highest angular momentum and densest sub-cluster in the SC2 simulation (the same as shown in Fig.~\ref{rot_comp}).
}\label{rot_evo}
\end{center}
\end{figure}

\begin{figure*}
\begin{center}
\includegraphics[scale=0.47]{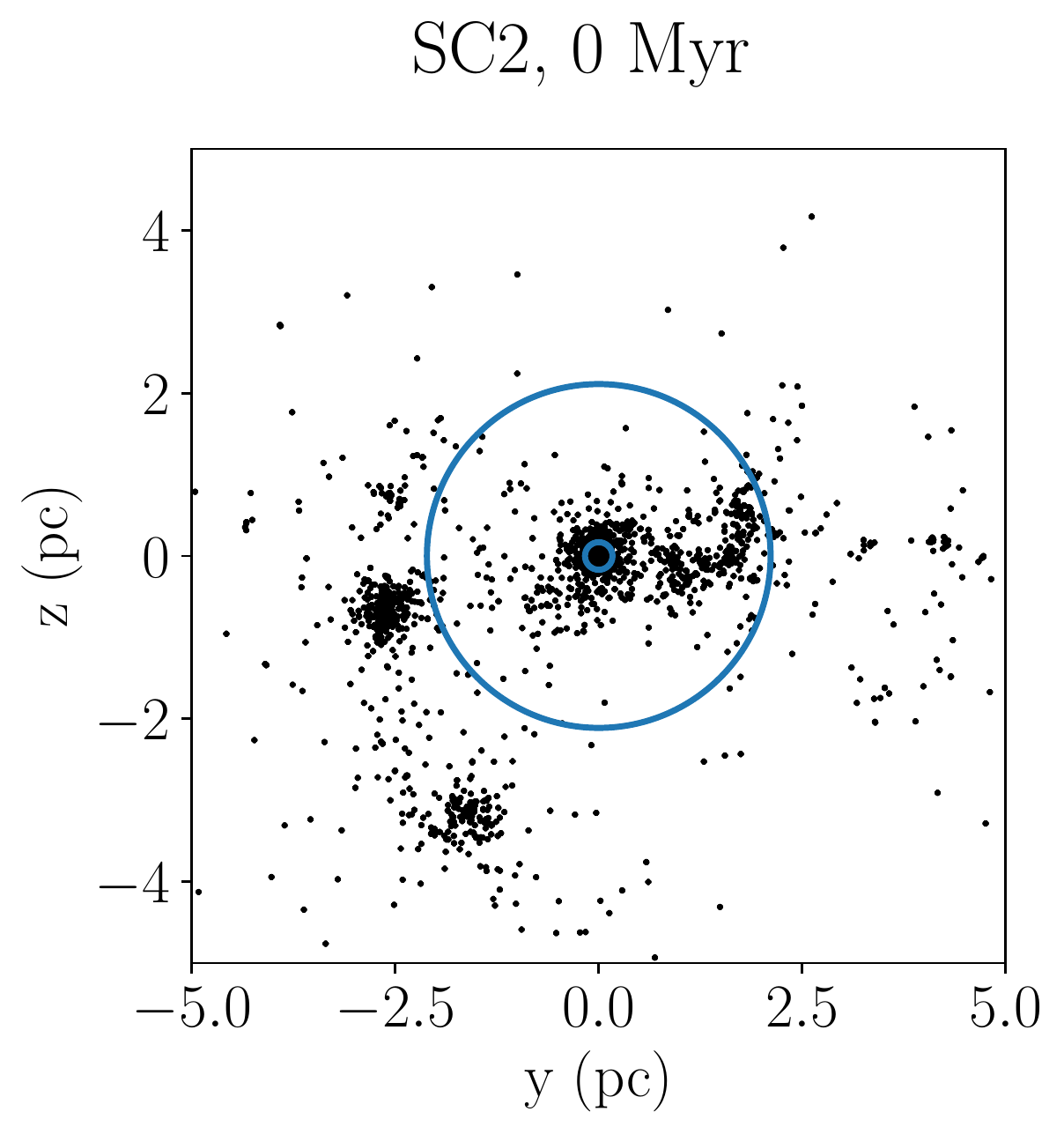}
\includegraphics[scale=0.47]{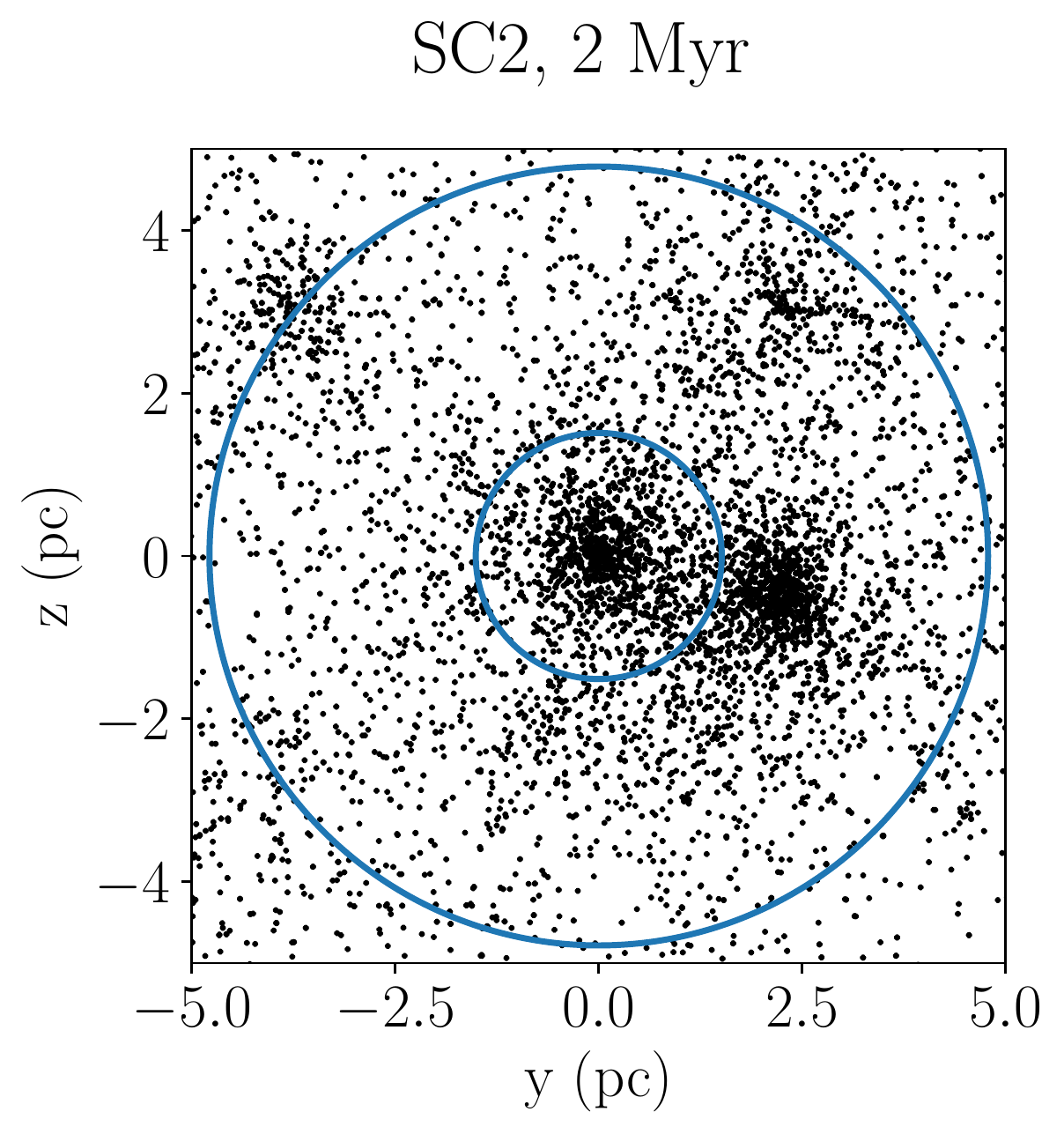}
\includegraphics[scale=0.47]{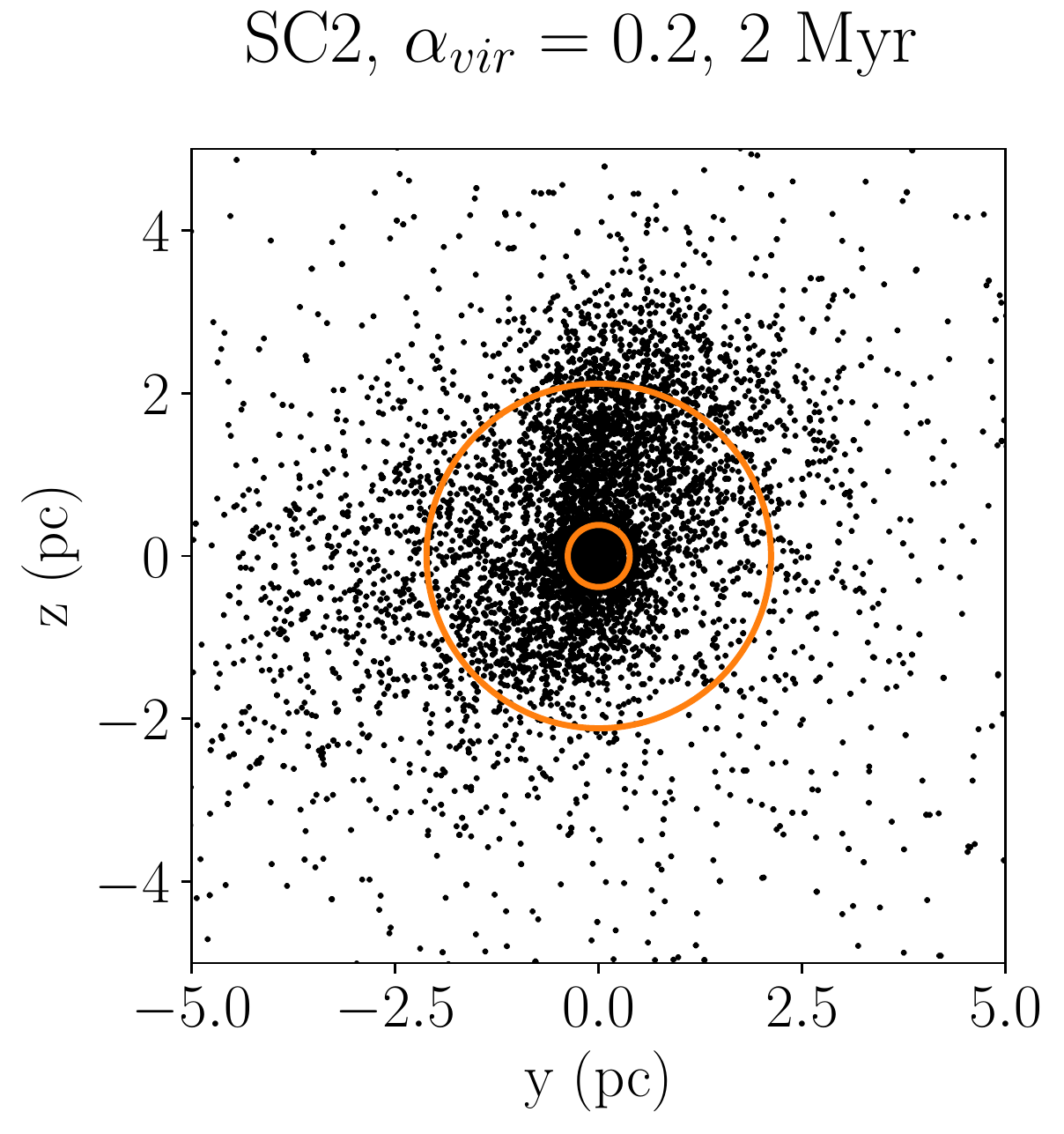}
\caption{Scatter plots of the distribution of stars for our SC2 cluster, at the beginning of the \textit{N}-body simulation (left-hand panel), after 2 Myr for the original virial ratio (central panel) and after 2 Myr for the reduced virial ratio run (right-hand panel). The outer and inner circles shows the 50\% and 10\% Lagrangian radii, respectively.
}\label{map_sc2}
\end{center}
\end{figure*}

\begin{figure}
\begin{center}
\includegraphics[scale=0.5]{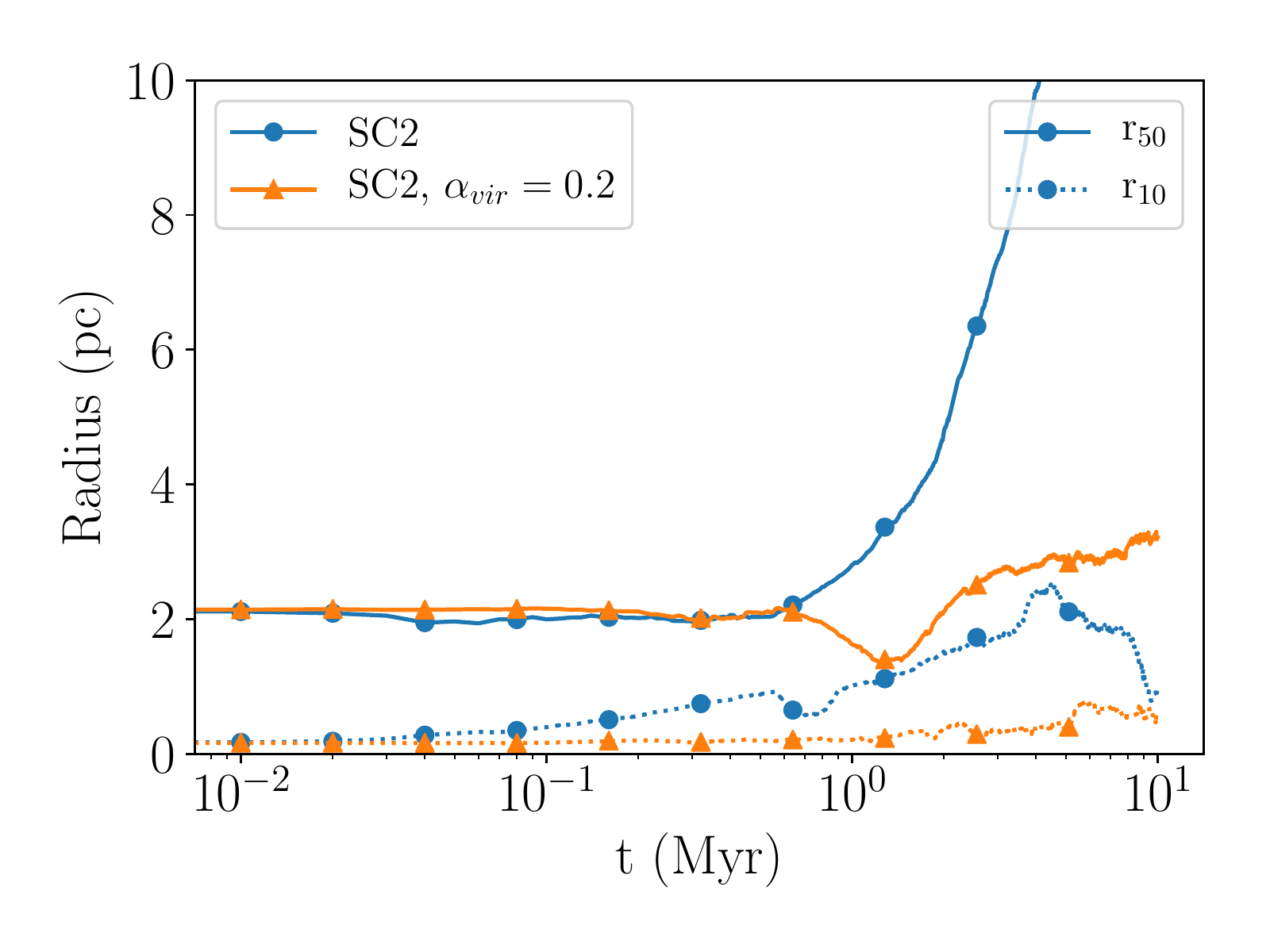}
\caption{Evolution of the 50\% (solid lines) and 10\% Lagrangian radius (dotted lines) for the \textit{N}-body rerun of the SC2 star cluster, with its original virial ratio (blue lines, circles) and with a reduced virial ratio of 0.3 (orange lines, triangles).
}\label{radii_vir}
\end{center}
\end{figure}

\begin{figure}
\begin{center}
\includegraphics[scale=0.5]{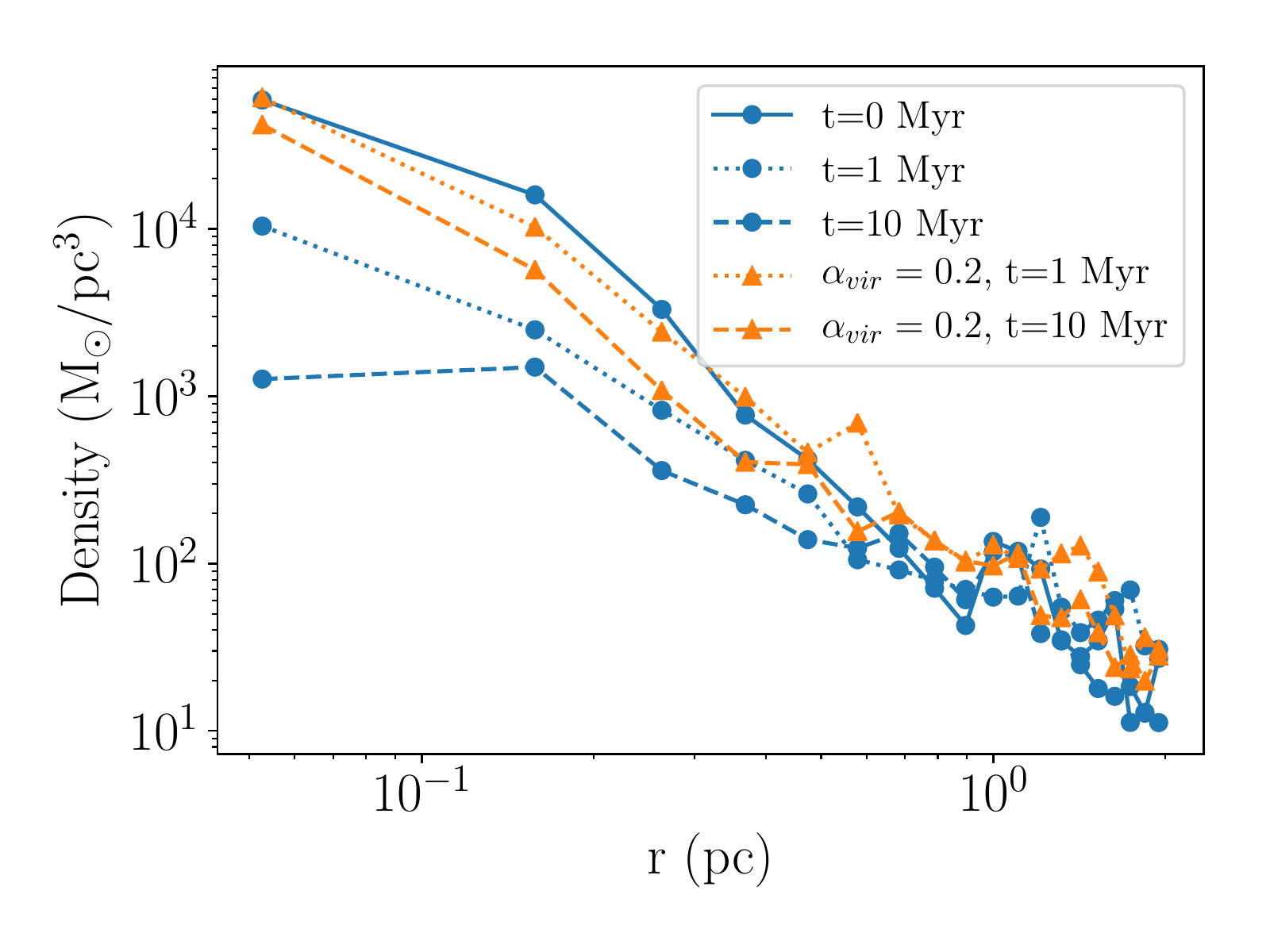}
\caption{Density profiles for the \textit{N}-body rerun of the SC2 star cluster, with its original virial ratio (blue line, circles) and with a reduced virial ratio of 0.3 (orange line, triangles), at the beginning of the \textit{N}-body evolution (solid line), after 1 Myr (dotted line) and after 10 Myr (dashed line).
}\label{den_vir}
\end{center}
\end{figure}

\begin{figure}
\begin{center}
\includegraphics[scale=0.5]{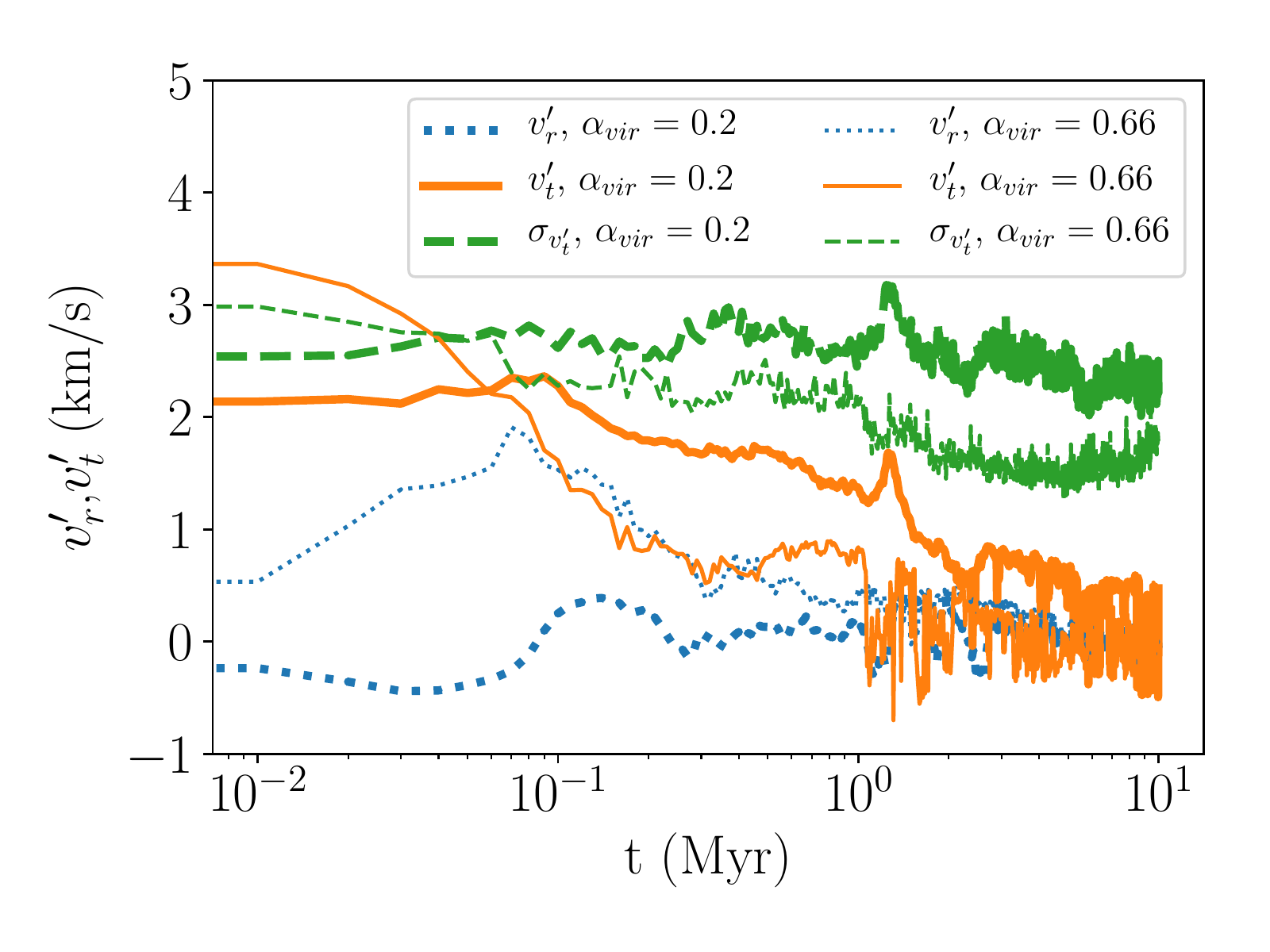}
\caption{Evolution of the tangential velocity (orange line), radial velocity (blue line) and tangential velocity dispersion (green line) for the highest angular momentum (and densest) sub-cluster in the SC2 simulation, with its original virial ratio (thin lines) and after reducing its virial ratio to 0.2 (thick lines).
}\label{rot_evo_vir}
\end{center}
\end{figure}

\section{Results}\label{results}

\subsection{Evolution of the cluster size}\label{sec_size}

Figure~\ref{radii_set} shows the early evolution ($<2$ Myr) of the 50\% Lagrangian radius $r_{50}$ (upper panel) and 10\% Lagrangian radius $r_{10}$ (lower panel). For the calculation of their evolution, these are the radii of spheres that are initially centered on the highest stellar density peak and at each time output they are centered on the highest stellar density peak within a distance of 0.2 pc from the one of the previous time outputs. This procedure was adopted to always center the whole cluster on the same sub-structure. Figure \ref{radii_set} shows that our clusters can have a different evolution of their size. For some clusters, $r_{50}$ starts increasing in about 0.1 Myr (see SC4, SC5 and SC8). In other cases, $r_{50}$ stays approximately constant for a longer time ($\gtrsim 0.5$ Myr) and then increases (see SC2, SC3, SC9 and SC10). SC1 shows an initial increase of $r_{50}$ after about 0.1 Myr, followed by a small decrease and then a further increase. For SC6, $r_{50}$ decreases until about 0.8-0.9 Myr and then it starts to increase, after.

The evolution of $r_{50}$ is due to two competing mechanisms. 
\begin{enumerate}
\item On one side, each sub-cluster tends to naturally dissolve. For our clusters, this is mostly due to the instantaneous gas removal, which leads to the dispersal of the outer stars of each sub-cluster. Although dissolution may also be caused by two-body internal relaxation, we found that this process is not the most important one in the early evolution of our systems. We estimated the central\footnote{For ``central'' we mean that each of the following quantities are calculated for the stars within the inner 0.2 pc of our clusters.} two-body relaxation timescale of our sub-clusters using the formula in \citet{Fujii14},

\begin{equation}\label{eq1}
    t_{2b,c}=\frac{0.065 \sigma^3_c}{G^2 \langle m\rangle \rho_c \ln\Lambda},
\end{equation}

where $\sigma^3_c$ is the central three-dimensional velocity dispersion, $G$ is the gravitational constant, $\langle m\rangle$ is the average particle mass, $\rho_c$ is the central mass density and $\ln\Lambda$ is the Coulomb logarithm, calculated as $\ln\gamma N$, with $N$ being the number of central particles and $\gamma=0.015$.
From this estimate, we found that $t_{2b,c}\gtrsim 20$ Myr for all simulations. On the contrary, from Fig. \ref{radii_set}, we expect the size of our (sub-)clusters to quickly react to the instantaneous gas removal, on timescales of the order of 0.1 Myr. Hence, two-body encounters are certainly happening in our dense sub-clusters, but they are not the main responsible for the evolution of their sizes. Furthermore, our clusters are highly sub-structured and are often composed of a large number of sub-clusters, as visible in Fig. \ref{map_stars}. Different sub-clusters might end up being unbound from the main, densest sub-cluster and move away from it, leading to a large increase of $r_{50}$ on short timescales. The dissolution of our systems is helped by the slightly super-virial state of our clusters. Our star clusters are slightly super-virial because the molecular clouds were initialized in a marginally bound state, i.e. with virial ratio equal to 1. Simulating initially sub-virial clouds would have led to a lower number of sub-clusters and to overall sub-virial star clusters \citep[as in the case, e.g., of][]{Kuznetsova15}.

\item On the other hand, if some sub-clusters are still bound to the densest one at the beginning of the \textit{N}-body simulation, they might need some time to merge in a single structure, keeping $r_{50}$ almost constant for a longer timescale. A major merger between two massive sub-clusters can even lead to a significant decrease of $r_{50}$, given our definition. These mergers happen on different timescales, depending on how bound the sub-clusters are to each other and on their initial distance. In another perspective, this process depends on whether (and how much) the total mass is distributed in one single or more major sub-clusters. Indeed, Fig. \ref{radii_set} shows a sort of correlation between the time at which the $r_{50}$ starts to increase and the initial size of the cluster, which reflects how the mass is initially distributed.
\end{enumerate}

An increase of the 10\% Lagrangian radius with time is also visible. Again, sub-structures reaching the center of our densest sub-clusters may cause small decreases of $r_{10}$. These dips are not necessarily visible in the evolution of $r_{50}$, since $r_{10}$ can be affected by small sub-structures reaching the center of the densest sub-cluster. In some cases, even if the mass of gas that is instantaneously removed is about twice the total mass of our star clusters (see section \ref{caveats}), $r_{10}$ is more slowly affected, since the gravitational field in the inner regions of our sub-structures was already dominated by sink particles at the time of the hydrodynamical simulations, which we decided to adopt as initial conditions for our \textit{N}-body simulations. This is visible in Fig. \ref{mass}, but also in Fig. \ref{density}, which shows the density profiles of sinks, post-joining/splitting stars and gas particles, for the main sub-cluster in the SC2 simulation. As also visible in this plot, at the time chosen to perform our joining/splitting, the amount of gas in the inner regions of the main sub-cluster is much lower than that of sinks/stars.

We can take the SC2 and SC6 models as two extreme cases of this complex interplay of mechanisms. 

SC2 is the second closest-to-virial model and starts with a main sub-cluster containing a large fraction of the total mass of the cluster, while the rest of the mass is mostly contained in a secondary major sub-cluster (see Figures \ref{map_stars} and \ref{var_split}). Nonetheless, its $r_{50}$ contains few minor sub-structures, which are bound to the main sub-cluster and are merging with the latter at around 0.6-0.7 Myr, reaching its very center. This minor mergers of strongly bound small sub-structures cause the dip observed in the evolution of its $r_{10}$ (see lower panel of Fig. \ref{radii_set}) and, at the same time, cause a small delay, compared to the other models, in the increase of its $r_{50}$, which stays constant for about 0.6-0.7 Myr (see upper panel of Fig. \ref{radii_set}). After this time, $r_{50}$ and (more slowly) $r_{10}$ have a clear increase, due to the escape of the rest of the cluster. 

SC6 has an almost opposite evolution. This is the most-supervirial model and it starts with several sub-structures being apart from each other (see Figures \ref{map_stars} and \ref{var_split}). This explains its large initial value of $r_{50}$ (see Table \ref{tab1} and the upper panel of Fig. \ref{radii_set}). During its evolution, some of these sub-structures get closer to densest one, leading to an initial monotonical decrease of $r_{50}$, getting to a minimum after about 0.8-0.9 Myr. In the case of SC6, however, this secondary sub-clusters never get to the very center of the main one. The main sub-cluster gets dispersed due to the instantaneous gas removal, a ``peripheral''  merger lead to the decrease in $r_{50}$, but never strongly affects $r_{10}$ (though some impact is visible at 0.8-0.9 Myr, in the lower panel of Fig. \ref{radii_set}). 

\subsection{Rotation}

In Figure \ref{rot_comp}, we show rotation scatter plots for the densest sub-cluster in the SC2 star cluster, before and after applying our joining/splitting. In order to produce these plots, we calculated the angular momentum of the sub-structure and we moved to a new frame of reference $x',y',z'$ where $z'$ is aligned with the direction of the angular momentum \citep[see also][]{Ballone20}. These figures show that the rotation of this sub-cluster is well preserved by our algorithm, even though the joining branch removed some stars in the densest region of the sink particle distribution.

Figure \ref{rot_evo} shows the time evolution of the average value of the tangential ($v'_{t}$) and radial ($v'_{r}$) components of the velocity in the angular momentum frame of reference, as well as the value of the dispersion of the tangential component ($\sigma_{v'_t}$).  In the first 0.1 Myr of evolution, the magnitude of the rotation (as measured by $v'_{t}$) decreases, while that of the radial motion increases, reaching a comparable value. After 0.1 Myr, both $v'_{t}$ and $v'_{r}$ decrease, reaching values $\lesssim 1$ km/s after 0.2 Myr. The value of $\sigma_{v'_t}$ also decreases in the first 0.2 Myr, reaching a constant value of $\gtrsim 2$ km/s for the further evolution.

This figure clearly shows that at $t\le{}0.1$ Myr the sub-structures have a rapid expansion, due to the the instantaneous gas removal (see also Section \ref{sec_size}), which leads to a fast decrease of the rotational velocity, simply explained by angular momentum conservation. With time, both the radial and tangential velocities tend to zero and the velocity dispersion settles to a value of around 2~km/s.

\subsection{Sub-virial cluster}

The lower right-hand panel of Fig. \ref{var_split} shows that all the clusters in our set are slightly super-virial. In order to test the impact of a lower energetic state of our clusters, we reran an \textit{N}-body simulation for the SC2 star cluster, after lowering its virial ratio to a value of $\alpha_{vir}=0.2$. This was done by simply rescaling all the velocities of the stars so to get a lower kinetic energy.

Figure \ref{map_sc2} shows the initial conditions (left) and the snapshot at 2 Myr of the super-virial (centre) and the sub-virial simulation ($\alpha_{vir}=0.2$, right). Initially, SC2 is composed of two main sub-clusters plus several smaller sub-clusters. The most massive sub-cluster is shown at the centre of the plot, while the second most-massive sub-cluster appears on the left of the densest one, in the left-hand panel of Fig. \ref{map_sc2}. 

In the super-virial case, the two main sub-structures of the star cluster are still clearly distinct after 2 Myr of evolution and keep on orbiting around a common center of mass (central panel of Fig.~\ref{map_sc2}): the second most massive cluster is now at the right of the most massive one.

In contrast, in the sub-virial case (right-hand panel of Fig.~\ref{map_sc2}), the two main sub-structures of the cluster merge after about 1 Myr from the beginning of the \textit{N}-body simulation, leading to a single (though still not fully relaxed) structure at 2 Myr. This explains the decrease of the 50\% Lagrangian radius of the sub-virial cluster in Fig.~\ref{radii_vir}. After this merger, the cluster keeps on relaxing, slowly increasing its $r_{50}$, which barely doubles after 10 Myr of dynamical evolution. Interestingly, the 10\% Lagrangian radius is more mildly affected by the merger. 

Figure~\ref{den_vir} shows the density profile of stars for the super-virial and sub-virial clusters. In the super-virial case, the gas removal (combined to some internal two-body relaxation) leads to a decrease of the density with time, over the inner 0.5 pc. In the sub-virial cluster, instead, the density distribution is less affected by gas removal and the cluster stays more bound for the first 10 Myr of its evolution. Again, this is also due to the smaller effect of gas removal on the more bound cluster and to the ``fueling'' of stars at about 1 Myr, through the merger with the second biggest sub-structure of the cluster.

The impact of the merger is also visible in the evolution of the rotation of the densest sub-cluster in the sub-virial case. In the first 0.1 Myr, the sub-cluster has an initially mild contraction, followed by an equally small re-expansion. The initial contraction leads to a small increase of the rotational velocity, followed by a decrease in the ``bouncing'' phase. In the first Myr, the rotation keeps on decreasing, due to some two-body relaxation. The merger of the two sub-structures is clearly visible both in the rotational velocity and in the velocity dispersion.  
Right after 1 Myr of evolution, the tangential velocity shows a small increase, due to the angular momentum carried by the second main sub-structure. At the same time, the merger also leads to an obvious increase in the velocity dispersion. After the merger, the rotation gets slowly ``erased'' by two-body relaxation in about 10 Myr. Indeed, we calculated with Eq. \ref{eq1} the central two-body relaxation at the beginning of this sub-virial case and, for this lower virial case, we found $t_{2b,c}\simeq 10$ Myr.

\section{Discussion}\label{disc}

As a first direct comparison with previous studies, we just checked the evolution of the size of our star clusters. Our results are widely in agreement with previous attempts of \textit{N}-body reruns of star clusters generated by hydrodynamical simulations of star formation \citep[e.g.,][]{Moeckel10, Parker13, Fujii16}. We showed that the half-mass radius of our clusters can have very little variation or increase by a factor up to 3 in the first 2 Myr of gas-free dynamical evolution. As discussed in Section \ref{sec_size}, this evolution is to be interpreted with a grain of salt, since different sub-structures of our extremely fractal star clusters can still hierarchically merge to form a more massive cluster within the first few Myr of our \textit{N}-body simulations.

We also showed that the initial energetic state of the cluster has a big impact on its evolution \citep[consistent with the results of previous studies, e.g.,][]{Baumgardt07, Kruijssen12, Farias15}. We looked at one of the clusters coming from our set of hydrodynamical simulation considering its original virial ratio, $\alpha_{vir}\approx0.66$, and a lower virial ratio of $\alpha_{vir}\approx0.2$. In the former case, the cluster more strongly feels the effect of its initial energetic state and of the instantaneous gas removal, more significantly increasing its size and becoming more diffuse. These super-virial clusters match the properties of star cluster associations and the smallest open clusters \citep[such as their radii and central densities; see, e.g.,][]{Fujii15b}. In contrast, the initially sub-virial cluster stays bound over 10 Myr, barely changing its density and size. Sub-virial clusters can experience a ``dry'' hierarchical assembly (i.e., different sub-clumps of the cluster can merge after this enters a gas-less phase), grow in mass and might look more similar to the observed young dense star clusters \citep[see, e.g.,][]{Fujii12, Fujii15b}.

The energetic state of the cluster and its hierarchical assembly are also crucial for the persistence of rotation. The sudden gas removal can immediately wash out the rotation acquired by the cluster in its embedded phase \citep[see][]{Lee16, Mapelli17, Ballone20, Chen20}. After gas removal, in fact, our sub-clusters can experience a short phase of expansion  (as visible in Fig. \ref{rot_evo}), particularly in the outer regions, led by a readjustment to the instantaneous lack of gas. Such expansion, at constant angular momentum, naturally reduces the angular velocity of the stars.
In contrast, rotation is erased on longer timescales, due to two-body relaxation, if the cluster is sufficiently sub-virial at the moment of gas expulsion, and it can even be fueled by angular momentum brought by mergers, as already discussed by \citet{Henault-Brunet12} for the 2 Myr old young star cluster R136 in the Large Magellanic Cloud \citep[as shown by][mergers can similarly explain the rotation features observed in globular clusters]{Mastrobuono-Battisti19}. 
Finally, we stress that the hierarchical assembly can happen independently of the initial virial ratio of the cluster. In fact, all our clusters are slightly super-virial (see Fig. \ref{var_split}), but mergers occur, in some cases, even after 1 Myr of gas-free evolution (see Fig. \ref{radii_set}).

\section{Summary and conclusions}

We presented a new method to obtain initial conditions for \textit{N}-body studies of young star clusters from hydrodynamical simulations of their formation from the collapse of massive molecular clouds. Our joining/splitting algorithm allows us to get a realistic mass spectrum of stars, which can be extremely important for the dynamical evolution of these systems. Here, we particularly focused on the early dynamical evolution of our young star clusters after they instantaneously lose, through supernova explosion, all the gas in which they were embedded. 
In particular, we showed that such initial gas-free evolution might be driven not only by two-body relaxation, but also by the sudden lack of the previously present gas potential and by a phase of dry mergers of different sub-structures of our highly fractal stellar distributions \citep[see][]{Ballone20}. We also showed that rotation of our (sub-)clusters can be rapidly erased by their expansion (if they are in a loose state) or it can persist for few Myr and even be fueled by mergers of sub-structures.

Our more realistic initial conditions can be used to study several other phenomena: e.g., in the near future we plan to investigate the evolution of the fractality of these clusters. The next fundamental step will consist of adding a realistic population of primordial binaries to our joining/splitting algorithm (Torniamenti et al., in preparation). Adding a realistic population of stellar binaries can also provide realistic initial conditions to study a plethora of different topics, such as the formation of blue stragglers \citep[e.g.,][]{Mapelli04,Mapelli06,Leigh07, Knigge09,Ferraro12,Antonini16,Portegies-Zwart19}, the build up of mass segregation \citep[e.g.,][]{Bonnell98,McMillan07,Allison09,Kuepper11}, the dynamical formation of X-ray binaries \citep[e.g.,][]{Linden10,Garofali12,Mapelli13a,Mapelli14,Goswami14,Johns-Mulia19}, and the properties  of binary compact objects in young star clusters  \citep[e.g.,][]{Ziosi14, Mapelli16, Banerjee17, Banerjee18a, Banerjee18b, Fujii17, DiCarlo19, DiCarlo20, Kumamoto19, Kumamoto20, Rastello20}.

\section*{Acknowledgements}

We thank the anonymous referee for their constructive criticism and Simon Portegies Zwart for useful suggestions.
AB, MM and SR acknowledge financial support by the European Research Council for the ERC Consolidator grant DEMOBLACK, under contract no. 770017. MS acknowledges funding from the European Union’s Horizon 2020 research and innovation programme under the Marie-Sklodowska-Curie grant agreement No. 794393. We acknowledge the CINECA award HP10BQM9PE under the ISCRA initiative and the CINECA-INFN agreement, for the availability of high performance computing resources and support.


\section*{Data availability}
The data underlying this article will be shared on reasonable request to the corresponding authors.



\bibliographystyle{mnras}
\bibliography{litdyn} 

\bsp	
\label{lastpage}
\end{document}